# Persistence of Strong Silica-Enriched Domains in the Earth's Lower Mantle

Maxim D. Ballmer[1,2], Christine Houser[1], John W. Hernlund[1], Renata M. Wentzcovitch[1,3,4] & Kei Hirose[1]

[1]*Earth-Life Science Institute, Tokyo Institute of Technology, Meguro, Tokyo, Japan.*

[2]*Institute of Geophysics, ETH Zurich, Zurich, Switzerland.*

[3]*Department of Applied Physics and Applied Mathematics, Columbia University, New York, USA.*

[4]*Department of Earth and Environmental Sciences, Columbia University, Lamont-Doherty Earth Observatory, Palisades NY, USA.*

**The composition of the lower mantle – comprising 56% of Earth's volume – remains poorly constrained. Among the major elements, Mg/Si ratios ranging from ∼0.9–1.1, such as in rocky solar-system building blocks (or chondrites), to ∼1.2–1.3, such as in upper-mantle rocks (or pyrolite), have been proposed. Geophysical evidence for subducted lithosphere deep in the mantle has been interpreted in terms of efficient mixing and thus homogeneous Mg/Si across most of the mantle. However, previous models did not consider the effects of variable Mg/Si on the viscosity and mixing efficiency of lower-mantle rocks. Here, we use geodynamic models to show that large-scale heterogeneity with viscosity variations of ∼20×, such as due to the dominance of intrinsically strong $(Mg,Fe)SiO_3$–bridgmanite in low-Mg/Si domains, are sufficient to prevent efficient mantle mixing, even on large scales. Models predict that intrinsically strong domains stabilize degree-two mantle-convection pat-**

**terns, and coherently persist at depths of ∼1,000–2,200 km up to the present-day, separated by relatively narrow up-/downwelling conduits of pyrolitic material. The stable manifestation of such "bridgmanite-enriched ancient mantle structures" (BEAMS) may reconcile the geographical fixity of deep-rooted mantle-upwelling centers, and fundamental geophysical changes near 1,000 km depth (e.g. in terms of seismic-tomography patterns, radial viscosity increase, lateral deflections of rising plumes and sinking slabs). Moreover, these ancient structures may provide a reservoir to host primordial geochemical signatures.**

State-of-the-art seismic-tomography models are difficult to reconcile with a mantle that is homogeneous (pyrolitic) on large length-scales. For example, most recently-subducted slabs flatten appearing to stagnate at either ∼660 km or ∼1,000 km depth[1]. Many mantle plumes are inferred to be deflected at similar depths[2,3]. In particular, deflections of mantle up-/downwellings in the uppermost lower mantle remain enigmatic. A viscosity increase near 1,000 km depth, consistent with geoid inversions, has been invoked to explain these observations[4,5]. However, there is no candidate phase transition to account for a sharp viscosity jump that could markedly affect mantle flow. Alternatively, compositional layering has been proposed[6], but the effects of coupled large-scale compositional and rheological heterogeneity on mantle dynamics remain poorly understood.

**Composition-induced viscosity variations in the lower mantle**

Lateral heterogeneity in lower-mantle composition can give rise to rheological contrasts. Heterogeneity involving $SiO_2$-enriched rocks has been put forward to balance the Earth's Si budget relative to the sun and chondrites, also given limitations to dissolve Si in the present-day outer core[7]. SiO$_2$-enriched rocks with CI-chondritic Mg/Si of ∼0.9–1.1 should host ∼87-97% (Mg,Fe)SiO$_3$−bridgmanite (Br) and only ∼0-10% (Mg,Fe)O-ferropericlase (Fp), in addition to a minor amount of Ca-perovskite (∼3%). In contrast, pyrolitic rocks with Mg/Si ∼1.2–1.3) contain only ∼75-80% Br and up to ∼17-23% Fp in the lower mantle. As the viscosity of Br is estimated to be ∼1,000 times greater than that of Fp[8], and rheological models for rocks consisting of two phases[9] predict highly non-linear variations in rock viscosity as the modal abundance of the weak phase varies betweeen 0%–30% (Suppl. Figure S5), any SiO$_2$-enriched rocks (with relatively low Mg/Si and Fp content) are significantly more viscous than pyrolite in the lower mantle.

Intrinsically viscous rocks are thought to resist entrainment by mantle convection and processing at spreading centers[10]. However, the style of mantle convection in the presence of intense rheological contrasts due to large-scale compositional heterogeneity has not yet been quantitatively explored. We perform a suite of two-dimensional numerical experiments initially including a layer of intrinsically stronger, and modestly denser SiO$_2$-enriched rock in the lower mantle than the pyrolitic SiO$_2$-depleted material in the upper mantle (see methods). Model viscosity depends on temperature and composition, but composition-dependent rheology is limited to the lower mantle, where Fp+Br are the dominant stable phases (Suppl. Info.). We assume that SiO$_2$-enriched material (pyroxenite in the upper mantle) undergoes partial melting at <125 km depth to leave a SiO$_2$-poor pyrolitic residue[11]. The precise viscosity and density contrasts that may be relevant for the Earth's lower mantle are poorly constrained; therefore, we vary both parameters systematically.

We observe two regimes in our numerical experiments. In regime A, both materials are read-ily mixed and the mantle becomes largely homogenized over time-scales shorter than the age of the Earth (Figure 1a-b). This regime occurs for relatively small viscosity contrasts between materials and is well-understood[12,13]. In regime B, we find instead that the intrinsically strong $SiO_2$-enriched material can avoid significant entrainment and mixing for model times greater than the age of the Earth (Figure 1c-f). A juxtaposition of both regimes is shown in Figure 1b-c as a comparison between the example case with moderate compositional viscosity and density contrasts (regime B), and reference case I with no such contrasts, but with a viscosity jump of factor $\lambda=8$ at 660 km depth (regime A). The viscosity jump is imposed to ensure comparable viscosity profiles and convective vigors between cases (Suppl. Info.).

**A new regime of mantle convection**

In the newly-described regime B, large-scale intrinsically strong $SiO_2$-enriched domains organize mantle-convection patterns. Initially, the upper-mantle pyrolitic material cools near the surface and soon sinks through the strong material in the lower mantle, thus forming relatively weak conduits. As the weaker material covers the core-mantle boundary and is heated, it becomes buoyant and rises upward through the strong layer to establish complementary upwelling conduits. Subsequently, the $SiO_2$-enriched material is encapsulated by the weaker pyrolite, which continues to circulate between the shallow and deepest mantle through the existing weak channels (Figure 1c-e). This encapsulation by weak material dramatically reduces stresses within strong domains. Therefore, strong domains – hereafter referred to as bridgmanite-enriched ancient mantle structures (BEAMS) – tend to avoid significant internal deformation, rather assuming slow coherent rotation.

The weaker pyrolitic material slowly but progressively entrains $SiO_2$-enriched material as it circulates around BEAMS. Conduits thus contain an assemblage of $SiO_2$-poor and $SiO_2$-enriched materials, the latter of which would manifest as a pyroxenite-like mafic rock in the upper mantle. Note however that the $SiO_2$-poor pyrolitic material itself may consist of a fine-scale mixture of ultramafic to mafic rocks with compositions ranging from harzburgite to mid-ocean-ridge basalt (MORB). Our models also predict the ingestion of some weak plumes into BEAMS, particularly during early stages, which become stretched out into spiral shapes that persist as fossil fragments. Nevertheless, for sufficiently large viscosity contrast BEAMS remain largely coherent and stabilize lower mantle convection patterns over billions of years (Figure 1c,f). Little material crosses over from one conveyor circuit to another, giving rise to long-lived chemically-isolated domains. This tendency for isolation of convection cells suggests a possible mechanism for producing global-scale variations in MORB geochemistry[14], and preserving primordial reservoirs[15].

Persistence of BEAMS for 4.6 Gyrs or longer is predicted for respective density and viscosity contrasts of ∼0.4% and >20 (Figure 2). These contrasts are consistent with the effects of variable Mg/Si (or Br-content) on lower mantle density and viscosity (see Suppl. Info.). Density contrasts of <0.25% or >1% demand somewhat greater viscosity contrasts for long-term persistence, because any related rising or sinking (respectively) of BEAMS enhances viscous entrainment.

In the 3D spherical-shell geometry of Earth's mantle, BEAMS likely assume somewhat more complex shapes than suggested by our 2D-Cartesian models. 3D-BEAMS should assume forms similar to donuts or rolls, minimizing internal deformation of high-viscosity domains (Suppl. Info.). Even though internal rotation of (donut-shaped) BEAMS may be difficult or even impossible in 3D, pyrolitic material would still circulate around BEAMS. Donut holes may accommodate upwelling centers (such as those beneath the Pacific and Africa), while downwelling curtains (such as those related to the subduction of Farallon and Tethys lithosphere) may occur between donuts/rolls. Such geometries are indeed suggested by maps of radially-averaged seismic velocities in the mid-mantle (Figure 3).

**Comparison with geophysical observations**

The BEAMS hypothesis can explain various seismic observations. We computed thermodynamic and thermoelastic properties for lower-mantle materials (see methods), and find that an average BEAMS mantle can match one-dimensional profiles such as PREM[16] (Suppl. Figure S2). Note however that one-dimensional seismic profiles alone are insufficient to discriminate between compositional models, particularly given current mineral-physics uncertainties[17–19] (Suppl. Info.). For example, a homogeneous pyrolitic mantle also provides an acceptable fit[20,21]. Nevertheless, the BEAMS model can further reconcile the fading of vertically-coherent fast anomalies (or subducted slabs) from tomography images in the mid-mantle[22–24] (Suppl. Info.). As BEAMS are intrinsically slightly faster than pyrolite due to higher Br contents, the seismic signal of slabs is predicted to fade relative to an average that is elevated by the presence of BEAMS (Suppl. Figure S7). Moreover, cluster analysis of shear-wave tomography models robustly requires three clusters at the inferred depths of BEAMS manifestation (∼1,000–2,200 km)[25], while only two clusters ("slow" and "fast") are required in the deep lower mantle[24]. The geographical distribution of the third "neutral" cluster indeed agrees well with that of BEAMS inferred from Figure 3. Finally, radial coherence of large-scale seismic structure at depths >1,000 km is unrelated to upper-mantle seismic structure or plate-tectonic features[4,26], and thus points to an independent mechanism for large-scale heterogeneity at depth.

In particular, two key seismic observations can be better explained in the context of the BEAMS mantle than in that of a homogeneous-pyrolitic mantle. A regionally manifested compositional viscosity jump across BEAMS tops offers a simple explanation for the stagnation of some slabs at ∼1,000 km depth[1], while other slabs readily sink through downwelling conduits at the same time[27]. Also, the location of stagnant slabs is consistent with the inferred geometry of BEAMS (Figure 3), and neutral clusters[25]. In turn, displacement of individual mantle plumes near 1,000 km depth[2,3] may be caused by circulation of mantle flow around BEAMS, and any related sub-horizontal "wind" in the upper mantle and transition zone.

The BEAMS hypothesis (Figure 4) further reconciles a range of other geophysical and geological constraints. For example, any mantle "wind" around BEAMS should be coupled to continental motions via cratonic keels, thereby supporting mountain building where it converges (i.e., above lower-mantle downwelling conduits such as across S-America and Asia)[28,29], and rifting where it diverges (i.e., above upwellings such as in E-Africa). Such coupling is reflected by quadrupole moments of plate-motion vectors, and quadrupole stability over ≥250 Myrs indicates that mantle-flow patterns persist through time[30], perhaps stabilized by BEAMS. Near the core-mantle boundary, mantle circulation is predicted to converge around upwelling conduits in order to focus the generation and assent of plumes[3,31] beneath Africa and the south-central Pacific. These zones of convergence would also be the natural place for any (Fe-rich) dense mantle material to pile up, consistent with seismic images of large low shear-velocity provinces (LLSVP)[25,32–34] (see Figure 4). The long-term geographical fixity of these piles and plume-upwelling zones[31] again requires a mechanism for stabilization of mantle-flow patterns such as BEAMS. Otherwise, piles would readily respond to changes in mantle flow[35]. Accordingly, BEAMS may constrain the shapes of LLSVP-piles above the core-mantle boundary without requiring a delicate balance between viscous drag and gravitational forces[36]. Furthermore, probabilistic inversions of the geoid indicate a maximum of mantle viscosity (or "viscosity hill") in the mid-mantle[4]. While a viscosity hill is not uniquely required by the data within uncertainties, it would indeed naturally arise from the manifestation of intrinsically strong BEAMS at about 1,000–2,200 km depth. We stress that the presence of BEAMS is not the only possible cause for any of these observations, but can provide a straightforward unified explanation, and thus should be thoroughly tested.

Future quantitative tests of the BEAMS hypothesis should involve systematic studies of seismic reflections and seismic anisotropy in the lower mantle. Our models predict that underside as well as out-of-plane reflections should preferentially occur near BEAMS margins with dominantly positive polarities. Whereas reflections and conversions of seismic waves have indeed been commonly observed in the uppermost lower mantle, e.g. near the expected tops of BEAMS[6,37], a systematic study that could map any large-scale compositional heterogeneity is lacking. The predicted circulation around BEAMS further implies vertically-fast seismic anisotropy within up-and downwelling conduits (due to lattice-preferred[38] or shape-preferred [9] orientation), as well as horizontally-fast anisotropy above and below BEAMS. The latter prediction is consistent with observations of anisotropy beneath the Tonga slab that stagnates at ∼1,000 km depth[1,39,40], but more detailed regional studies of mid-mantle anisotropy are needed.

**Geochemical implications**

The geochemical implications of the BEAMS hypothesis depend on the origin scenario. An initial global lower-mantle $SiO_2$-enrichment compatible with our model starting conditions could arise due to (1) incomplete equilibration of the proto-mantle during multi-stage core formation[41], (2) fractionation during magma-ocean crystallization[42], and/or (3) continental extraction that leaves the shallow pyrolitic domain as a "depleted MORB mantle" residue. If BEAMS formed within ∼100 Myrs after Earth's formation (scenarios 1 and/or 2), then they would be viable candidates for hosting primordial noble-gas reservoirs[43,44] as well as primordial $^{182}W$[45], because BEAMS material is never processed through the shallow upper mantle. Note that at least in scenario (2) BEAMS would moreover be better candidates to host primordial geochemical signatures (such as e.g. FOZO[46]) than LLSVPs, because they would be relatively depleted in incompatible elements[47]. The predicted dynamical behavior of mostly stable BEAMS with gradual entrainment along margins provides the conditions for primordial reservoirs to be preserved in a vigorously convecting mantle, but also be sampled by hotspot lavas at the same time, along with recycled geochemical components[46–48]. In contrast to small-scale blobs that have previously been invoked to host primitive material[10,49], BEAMS can provide a large-scale coherent primordial reservoir of up to 10%~15% of the mantle's mass (Suppl. Info.). Such large-scale heterogeneity may balance Earth's bulk composition, e.g. bringing it closer to solar-chondritic Mg/Si-ratios.

**Methods**

We here describe the methodology of geodynamic models, as well as of the computation of thermodynamic and thermoelastic properties. For figures and more detailed discussion, we refer the reader to the main text as well as the Suppl. Information.

**Numerical mantle-convection models.** In order to study thermochemical convection of the mantle, we used an advanced version of mantle-convection code CitcomCU[52,53]. On the finite-element mesh, we solved the conservation equations of mass, momentum and energy applying the Boussinesq approximation. Composition is tracked using passive particles (or "tracers"). The model box is 2,900 km deep and 17,400 km wide. The vertical resolution of the model varies between 16.5 km and ∼18.7 km due to mesh refinement in the upper mantle. Horizontal resolution is 17 km. Initial conditions involve a difference in composition between the upper and lower mantles. In the upper mantle, tracer values are set to a compositional index of zero, representing $SiO_2$-poor mantle material similar to pyrolite. In the lower mantle, tracer values are randomly set to a compositional index of $0.95 \pm 0.05$, representing $(Mg,Fe)SiO_3$-rich (or $SiO_2$-rich) mantle material (Suppl. Figure S1). Random compositional noise of $\pm 0.05$ is added in the lower mantle in order to seed small non-diffusive perturbations that help to break the strong deep layer. Initial potential temperatures are 2,000 °C in the mid-mantle with thermal boundary layers at the top and bottom (calculated from 80-Myr halfspace cooling profiles), plus a small random thermal noise. Boundary conditions involve potential temperatures of $T_{surf}$ = 0 °C and $T_{CMB}$ = 3000 °C at the top and bottom, respectively, as well as free-slip velocity conditions on all sides. The applied $T_{CMB}$ is well in the range of estimates[54,55] (note that the adiabat needs to be added to $T_{CMB}$ for proper comparison with estimates of "real" CMB temperature).

Distinct physical properties are assigned to the two materials. $SiO_2$-rich material is denser (by $\Delta\rho$) and stiffer (by a factor of $\Phi$) than peridotitic material. While the density difference is applied everywhere in the mantle, the viscosity contrast is only applied in the lower mantle. This parameterization is motivated by the limitation of the stability of Br and Fp (i.e., to lower-mantle pressures), the presence of which in variable proportions between the materials is envisioned to account for the viscosity contrast (see Suppl. Info. and main text). Additionally, we prescribed that all "$SiO_2$-enriched" tracers, which enter the shallowest part of the mantle (i.e. at depths <125 km), are immediately turned into "pyrolitic tracers" (i.e., tracer values are set to zero), assuming that $SiO_2$-rich material undergoes melting to become relatively enriched in MgO. Such a depth of melting for $(Mg,Fe)SiO_3$-rich rocks is supported experimentally[11].

In our geodynamic models, we applied a Newtonian rheology with moderate temperature dependence of viscosity, and no depth dependence. Viscosity varies by six orders of magnitude over the full thermal range of $T_{CMB} - T_{surf}$, but a cutoff is applied at four orders of magnitude in the stiff thermal boundary layer at the top (see Figure 1e in the main text) in order to ensure numerical stability. Depth-dependency of thermal expansivity is accounted for (according to ref. [6]). For all other parameters, see Suppl. Table 2.

In order to systematically study the effects of intrinsic variations in density and viscosity on mantle flow, we performed a systematic parameter search by varying $\Delta\rho$ and $\Phi$. For a list of all cases, see Suppl. Table 3. $\Delta\rho$ is varied in the range of 0 and 65 kg/m$^3$ (i.e., 0%-1.444%), and $\Phi$ in the range of 3.136 to 249.1. We explored this parameter space by running 26 simulations with no imposed viscosity jump at 660 km depth (i.e., $\lambda = 1$). A regionally variable viscosity jump at 660 km depth self-consistently arises from our treatment of compositional rheology: as compositional rheology is restricted to the lower mantle (see above), a viscosity jump arises wherever compositional index >0. We also explored three reference cases $\lambda > 1$. For a detailed description and discussion of these reference cases, as well as for the post-processing and analysis of numerical-model predictions, see Suppl. Information.

**Computation of seismic velocities and densities.** One-dimensional seismic-velocity and density profiles are calculated for comparison with PREM[16] (Suppl. Figure S2). For this calculation, we used thermodynamic and thermoelastic properties of $Mg_{1-x}Fe_xSiO_3$ bridgmanite (Br) and $Mg_{1-y}Fe_yO$ ferropericlase (Fp) as previously computed by refs. [56] and [57,58], respectively, for iron numbers $x = 0$ and $x = 0.125$, as well as $y = 0$ and $y = 0.1875$. For all other $x$ and $y$ values, physical properties have been linearly interpolated. For $CaSiO_3$ perovskite, thermoelastic properties calculated by Kawai and Tsuchiya[59] were reproduced within the Mie-Debye-Grüneisen formalism as outlined by Stixrude and Lithgow-Bertelloni[60] using density functional theory (DFT) within the local density approximation (LDA) that is augmented by the Hubbard U (LDA+U). These thermoelastic properties were calculated self-consistently for Fe, Si, and O along with psuedopotentials for Mg. Details of the LDA and LDA+U calculations are reported in refs. [56–59].

We considered mixtures in the $SiO_2$–$MgO$–$CaO$–$FeO$ oxide space for aggregates with harzburgitic[61], pyrolytic[62], and perovskititic (i.e. pure Br) compositions[63]. For the specific oxide compositions of these aggregates, see Suppl. Table 1. Perovskititic compositions have been computed by in-crementally removing MgO from pyrolite. Note that calculations do not incorporate the effects of Al$_2$O$_3$. In the adjusted compositions, the number of moles of Al$_2$O$_3$ have been equally distributed between MgO and SiO$_2$. The iron partitioning coefficient, $K_D$, between Fp and Br was kept constant at 0.5[64]. Density and seismic-velocity profiles for these end-member compositions are shown in Supplementary Figure S3.

To compute these profiles, we used the self-consistent geotherms shown in Supplementary Figure S4. Moduli and densities for each of the minerals were interpolated along the calculated geotherms; physical properties of mineral assemblages have been obtained using the Voigt-Reuss-Hill (VRH) average. To calculate the adiabatic geotherms, the following equation has been integrated to solve for T(P)[65], where the aggregate quantities are the molar volume, the thermal expansion coefficient, and the isobaric specific heat of aggregates: $(\partial T/\partial P)_S = \alpha_{agg} V_{agg} T / C_{p_{agg}}$. In these calculations, the temperature at the top of the lower mantle (23 GPa) is anchored at 1873 K, as constrained by the post-spinel transition[66].

Finally, to compute density and seismic-velocity profiles for the BEAMS mantle, we used an idealized average composition of the lower mantle. Inspired by our numerical-model predictions, we assumed that 50% of the lower mantle is composed of perovskitite (i.e. pure Br), and 25% is composed of each cold and warm harzburgite (downwellings and upwellings, respectively). The relevant adiabats of these components are shown in Supplementary Figure S4.

**Method references.** Reference numbers 51-65 (see reference list below).

**References**

1. Fukao, Y. & Obayashi, M. Subducted slabs stagnant above, penetrating through, and trapped below the 660 km discontinuity. *J. Geophys. Res.* **118**, 5920–5938 (2013).

2. Rickers, F., Fichtner, A. & Trampert, J. The Iceland–Jan Mayen plume system and its impact on mantle dynamics in the North Atlantic region: Evidence from full-waveform inversion. *Earth and Planetary Science Letters* **367**, 39–51 (2013).

3. French, S. W. & Romanowicz, B. Broad plumes rooted at the base of the Earth's mantle beneath major hotspots. *Nature* **525**, 95–99 (2015).

4. Rudolph, M., Lekic, V. & Lithgow-Bertelloni, C. Viscosity jump in Earth's mid-mantle. *Science* **350**, 1349–1352 (2015).

5. Marquardt, H. & Miyagi, L. Slab stagnation in the shallow lower mantle linked to an increase in mantle viscosity. *Nature Geoscience* **8**, 311–314 (2015).

6. Ballmer, M. D., Schmerr, N. C., Nakagawa, T. & Ritsema, J. Compositional mantle layering revealed by slab stagnation at ∼1000-km depth. *Science Advances* **1**, doi:10.1126/sciadv.1500815 (2015).

7. Badro, J., Côté, A. S. & Brodholt, J. P. A seismologically consistent compositional model of Earth's core. *Proceedings of the National Academy of Sciences* **111**, 7542–7545 (2014).

8. Yamazaki, D. & Karato, S.-I. Some mineral physics constraints on the rheology and geothermal structure of Earth's lower mantle. *Amer. Mineralogist* **86**, 385–391 (2001).

9. Girard, J., Amulele, G., Farla, R., Mohiuddin, A. & Karato, S.-I. Shear deformation of bridgmanite and magnesiowüstite aggregates at lower mantle conditions. *Science* **351**, 144–147 (2016).

10. Manga, M. Low-viscosity mantle blobs are sampled preferentially at regions of surface divergence and stirred rapidly into the mantle. *Physics of the Earth and Planetary Interiors* **180**, 104–107 (2010).

11. Pertermann, M. & Hirschmann, M. M. Partial melting experiments on a MORB-like pyroxenite between 2 and 3 GPa: Constraints on the presence of pyroxenite in basalt source regions from solidus location and melting rate. *Journal of Geophysical Research: Solid Earth* **108**, 2125, doi:10.1029/2000JB000118 (2003).

12. Kellogg, L. Chaotic mixing in the Earth's mantle. In Dmowska, R. & Saltzman, B. (eds.) *Advances in Geophysics*, vol. 34, 1–33 (Academic Press, 1993).

13. Coltice, N. & Ricard, Y. Geochemical observations and one layer mantle convection. *Earth Planet. Sci. Letter.* **74**, 125–137 (1999).

14. Dupré, B. & Allègre, C. J. Pb-Sr isotope variation in Indian Ocean basalts and mixing phenomena. *Nature* **303**, 142–146 (1983).

15. Schubert, G. & Spohn, T. Two-layer mantle convection and the depletion of radioactive elements in the lower mantle. *Geophys. Res. Lett.* **8**, 951–954 (1981).

16. Dziewonski, A. & Anderson, D. Preliminary reference Earth model. *Phys. Earth Planet. Inter.* **25**, 297–356 (1981).

17. Murakami, M., Ohishi, Y., Hirao, N. & Hirose, K. A perovskitic lower mantle inferred from high–pressure, high–temperature sound velocity data. *Nature* **485**, 90–95 (2012).

18. Ismailova, L. *et al.* Stability of Fe, Al-bearing bridgmanite in the lower mantle and synthesis of pure Fe-bridgmanite. *Science Advances* **2**, doi:10.1126/sciadv.1600427 (2016).

19. Matas, J., Bass, J., Ricard, Y., Mattern, E. & Bukowinski, M. On the bulk composition of the lower mantle: predictions and limitations from generalized inversion of radial seismic profiles. *Geophysical Journal International* **170**, 764–780 (2007).

20. Wang, X., Tsuchiya, T. & Hase, A. Computational support for a pyrolitic lower mantle containing ferric iron. *Nature Geoscience* **8**, 556–559 (2015).

21. Hyung, E., Huang, S., Petaev, M. I. & Jacobsen, S. B. Is the mantle chemically stratified? Insights from sound velocity modeling and isotope evolution of an early magma ocean. *Earth and Planetary Science Letters* **440**, 158–168 (2016).

22. Hernlund, J. & Houser, C. On the distribution of seismic velocities in Earth's deep mantle. *Earth Planet. Sci. Lett.* **265**, 423–437 (2008).

23. Houser, C. & Williams, Q. The relative wavelengths of fast and slow velocity anomalies in the lower mantle: Contrary to the expectations of dynamics? *Phys. Earth Planet. Int.* **176**, 187–197 (2009).

24. Lekic, V., Cottaar, S., Dziewonski, A. & Romanowicz, B. Cluster analysis of global lower mantle tomography: A new class of structure and implications for chemical heterogeneity. *Earth Planet. Sci. Lett.* **357-358**, 68–77 (2012).

25. Cottaar, S. & Lekic, V. Morphology of seismically slow lower-mantle structures. *Geophysical Journal International* **207**, 1122–1136 (2016).

26. Dziewonski, A., Lekic, V. & Romanowicz, B. Mantle anchor structure: An argument for bottom up tectonics. *Earth and Planetary Science Letters* **299**, 69–79 (2010).

27. Van Der Meer, D. G., Spakman, W., Van Hinsbergen, D. J., Amaru, M. L. & Torsvik, T. H. Towards absolute plate motions constrained by lower-mantle slab remnants. *Nature Geoscience* **3**, 36–40 (2010).

28. Becker, T. & Faccenna, C. Mantle conveyor beneath the Tethyan collisional belt. *Earth and Planetary Science Letters* **310**, 453–461 (2011).

29. Faccenna, C., Becker, T., Conrad, C. & Husson, L. Mountain building and mantle dynamics. *Tectonics* **32**, 80–93 (2013).

30. Conrad, C. P., Steinberger, B. & Torsvik, T. H. Stability of active mantle upwelling revealed by net characteristics of plate tectonics. *Nature* **498**, 479–482 (2013).

31. Torsvik, T. H., Burke, K., Steinberger, B., Webb, S. J. & Ashwal, L. D. Diamonds sampled by plumes from the core-mantle boundary. *Nature* **466**, 352–355 (2010).

32. Garnero, E. J. & McNamara, A. K. Structure and dynamics of Earth's lower mantle. *Science* **320**, 626–628 (2008).

33. Deschamps, F., Cobden, L. & Tackley, P. J. The primitive nature of large low shear-wave velocity provinces. *Earth and Planetary Science Letters* **349**, 198–208 (2012).

34. Ballmer, M. D., Schumacher, L., Lekic, V., Thomas, C. & Ito, G. Compositional layering within the large low shear-wave velocity provinces in the lower mantle. *Geochemistry, Geophysics, Geosystems* **17**, doi:10.1002/2016GC006605 (2016).

35. McNamara, A. K. & Zhong, S. Thermochemical structures beneath Africa and the Pacific Ocean. *Nature* **437**, 1136–1139 (2005).

36. Tan, E. & Gurnis, M. Compressible thermochemical convection and application to lower mantle structures. *Journal of Geophysical Research: Solid Earth* **112**, doi:10.1029/2006JB004505 (2007).

37. Jenkins, J., Deuss, A. & Cottaar, S. Converted phases from sharp 1000 km depth mid-mantle heterogeneity beneath western europe. *Earth and Planetary Science Letters* **459**, 196–207 (2016).

38. Tsujino, N. *et al.* Mantle dynamics inferred from the crystallographic preferred orientation of bridgmanite. *Nature* **539**, 81–84 (2016).

39. Wookey, J. & Kendall, J.-M. Evidence of midmantle anisotropy from shear wave splitting and the influence of shear-coupled P waves. *Journal of Geophysical Research: Solid Earth* **109**, doi:10.1029/2003JB002871 (2004).

40. Chang, S.-J., Ferreira, A. M. & Faccenda, M. Upper-and mid-mantle interaction between the Samoan plume and the Tonga-Kermadec slabs. *Nature communications* **7**, doi:10.1038/ncomms10799 (2016).

41. Kaminski, E. & Javoy, M. A two-stage scenario for the formation of the Earth's mantle and core. *Earth and Planetary Science Letters* **365**, 97–107 (2013).

42. Elkins-Tanton, L. Linked magma ocean solidification and atmospheric growth for Earth and Mars. *Earth and Planetary Science Letters* **271**, 181–191 (2008).

43. Mukhopadhyay, S. Early differentiation and volatile accretion recorded in deep-mantle neon and xenon. *Nature* **486**, 101–104 (2012).

44. Caracausi, A., Avice, G., Burnard, P. G., Füri, E. & Marty, B. Chondritic xenon in the Earth's mantle. *Nature* **533**, 82–85 (2016).

45. Rizo, H. *et al.* Preservation of Earth-forming events in the tungsten isotopic composition of modern flood basalts. *Science* **352**, 809–812 (2016).

46. Hart, S., Hauri, E. *et al.* Mantle plumes and entrainment: Isotopic evidence. *Science* **256**, 517–520 (1992).

47. White, W. Isotopes, DUPAL, LLSVPs, and Anekantavada. *Chemical Geology* **419**, 10–28 (2015).

48. Garapić, G., Mallik, A., Dasgupta, R. & Jackson, M. G. Oceanic lavas sampling the high-3He/4He mantle reservoir: Primitive, depleted, or re-enriched? *American Mineralogist* **100**, 2066–2081 (2015).

49. Becker, T. W., Kellogg, J. B. & O'Connell, R. J. Thermal constraints on the survival of primitive blobs in the lower mantle. *Earth and Planetary Science Letters* **171**, 351–365 (1999).

50. Parmentier, E. M., Turcotte, D. L. & Torrance, K. E. Studies of finite amplitude non-Newtonian thermal convection with application to convection in the Earth's mantle. *Journal of Geophysical Research* **81**, 1839–1846 (1976).

51. Houser, C., Masters, G., Shearer, P. & Laske, G. Shear and compressional velocity models of the mantle from cluster analysis of long-period waveforms. *Geophys. J. Int.* **174**, 195–212 (2008).

52. Moresi, L., Zhong, S. & Gurnis, M. The accuracy of finite element solutions of Stokes's flow with strongly varying viscosity. *Physics of the Earth and Planetary Interiors* **97**, 83–94 (1996).

53. Ballmer, M., Van Hunen, J., Ito, G., Bianco, T. & Tackley, P. Intraplate volcanism with complex age-distance patterns: A case for small-scale sublithospheric convection. *Geochemistry, Geophysics, Geosystems* **10**, doi:10.1029/2009GC002386 (2009).

54. Lay, T., Hernlund, J. & Buffett, B. A. Core–mantle boundary heat flow. *Nature Geoscience* **1**, 25–32 (2008).

55. Nomura, R. *et al.* Low core-mantle boundary temperature inferred from the solidus of pyrolite. *Science* **343**, 522–525 (2014).

56. Shukla, G. *et al.* Thermoelasticity of $Fe^{2+}$-bearing bridgmanite. *Geophys. Res. Lett.* **42**, 1741–1749 (2015).

57. Wu, Z., Justo, J., da Silva, C., de Gironcoli, S. & Wentzcovitch, R. M. Anomalous thermodynamic properties in ferropericlase throughout its spin crossover transition. *Physical Review B* **80**, doi:10.1103/PhysRevB.80.014409 (2009).

58. Wu, Z., Justo, J. & Wentzcovitch, R. Elastic anomalies in a spin-crossover system: Ferropericlase at lower mantle conditions. *Phys. Rev. Lett.* **110**, doi:10.1103/PhysRevLett.110.228501 (2013).

59. Kawai, K. & Tsuchiya, T. P-V-T equation of state of cubic $CaSiO_3$ perovskite from first-principles computation. *J. Geophys. Res.* **119**, 2801–2809 (2014).

60. Stixrude, L. & Lithgow-Bertelloni, C. Thermodynamics of mantle minerals -I. Physical properties. *Geophys. J. Int.* **162**, 610–632 (2005).

61. Baker, M. & Beckett, J. The origin of abyssal peridotites: A reinterpretation of constraints based on primary bulk compositions. *Earth Planet. Sci. Lett.* **171**, 49–61 (1999).

62. McDonough, W. & Sun, S. The composition of the Earth. *Chem. Geo.* **120**, 223–253 (1995).

63. Williams, Q. & Knittle, E. The uncertain major element bulk composition of Earth's mantle. *Earth's Deep Mantle: Structure, Composition, and Evolution. Geophysical Monograph Series* **160**, 187–199 (2005).

64. Irifune, T. *et al.* Iron partitioning and density changes of pyrolite in Earth's lower mantle. *Science* **327**, 193–195 (2010).

65. Valencia-Dardona, J., Shukla, G., Wu., Z., Yuen, D. & Wentzcovitch, R. Impact of spin crossover in ferropericlase on the lower mantle adiabat. *Geophys. Res. Lett.* in review (2016).

66. Irifune, T. *et al.* The postspinel phase boundary in $Mg_2SiO_4$ determined by in situ X-ray diffraction. *Science* **279**, 1698–1700 (1998).

**Acknowledgements.** The authors thank Frédéric Deschamps and an anonymous reviewer for their comments that helped to improve the manuscript. They are further grateful to Adam Dziewonski for discussions on the seismic structure and dynamics of the mantle. Calculations have been performed on akua, the in-house cluster of the Department of Geology&Geophysics, Univ. Hawaii. The numerical tools used here, as well as model input and output, are available upon request. M.D.B., C.H., J.W.H., and K.H. were supported by the WPI-funded Earth-Life Science Institute at Tokyo Institute of Technology. C.H., J.W.H., and K.H. received further support through MEXT KAKENHI grant numbers 15H05832 and 16H06285. R.M.W. was funded through NSF grants EAR-1319361 and EAR1348066.

**Author Contributions.** M.D.B., C.H., and J.W.H. wrote the manuscript and composed the figures. M.D.B. performed and analyzed geodynamic models. C.H. and R.M.W. computed seismic velocities in the lower mantle. J.W.H. and K.H. analyzed the influence of composition on density and viscosity. All authors contributed to the BEAMS hypothesis, and the design of the study.

**Competing Financial Interests.** The authors declare that they have no competing financial interests.

**Corresponding authors.** Correspondence and requests for materials should be addressed to M.D.B..

**Figure 1** Numerical-model results for (A-B) reference case I, and the (C-F) example case after (A-E) 4.6 Gyrs and (F) 10 Gyrs model time. (B,C,F) Snapshots of composition with isotherms (spaced 450 K). (A,D) Snapshots of temperature with compositional contour that marks small-scale heterogeneity in (A), and large-scale BEAMS in (D). This difference in mantle-mixing efficiency between cases highlights the role of compositional rheology, given that both cases have similar Nusselt numbers $Nu$ (Suppl. Table S3), i.e. a criterion for convective vigor[50]). (E) Snapshot of viscosity shows that BEAMS are more viscous than upwelling and downwelling conduits. Also see Suppl. Movies S1-S4.

**Figure 2** Summary of numerical-model results. Regime map of all cases (Suppl. Table 3) shows that compositional-viscosity contrasts of ∼1.5 orders of magnitude and small-to-moderate compositional density contrasts are required for long-term persistence of $SiO_2$-enriched material (blue squares). This conclusion is independent of whether all cases, or the subset of cases with $10 \leq Nu \leq 11$ (yellow highlighted) are considered. In reference cases I/II and III (circles), a global viscosity jump at 660 km depth of factor $\lambda=8$ and $\lambda=2.5$, respectively, is imposed to ensure that $Nu$ is comparable to $Nu$ of the example case (Figure 1c-f), which is marked by a white cross.

**Figure 3** Map with possible distributions of BEAMS in the Earth's lower mantle. Colors show mid-mantle shear-velocity anomalies[51], radially averaged as annotated. As LLSVPs are primarily confined to 2,300-2,891 km depth[22,24], they do not dominate the radial average shown here. Note that the blue fast anomalies (downwelling conduits: "1","3"), are

∼2× weaker than the red slow anomalies (upwelling conduits: "2","4") (Suppl. Figure S7). BEAMS likely occupy the volume between conduits (dashed outlines); arrows mark the sense of associated upper-mantle flow. Stagnant slabs[1] ("S") should overlie BEAMS, guiding our assessment of BEAMS distributions, which well agree with cluster analysis of seismic-tomography models[25].

**Figure 4** Illustration of the BEAMS hypothesis. BEAMS (light grey) are stable high-viscosity structures that reside in Earth's lower mantle, while streaks of pyrolitic-harzburgitic rocks (light blue/green) and basalt (dark blue/green) circulate between the shallow and deep mantle through rheologically weak channels. BEAMS can coexist with, and stabilize the LLSVPs in the lowermost ∼500 km of the mantle (yellow), which are interpreted as intrinsically-dense (Fe-rich) piles[32,33,35] and plume-generation zones[31].

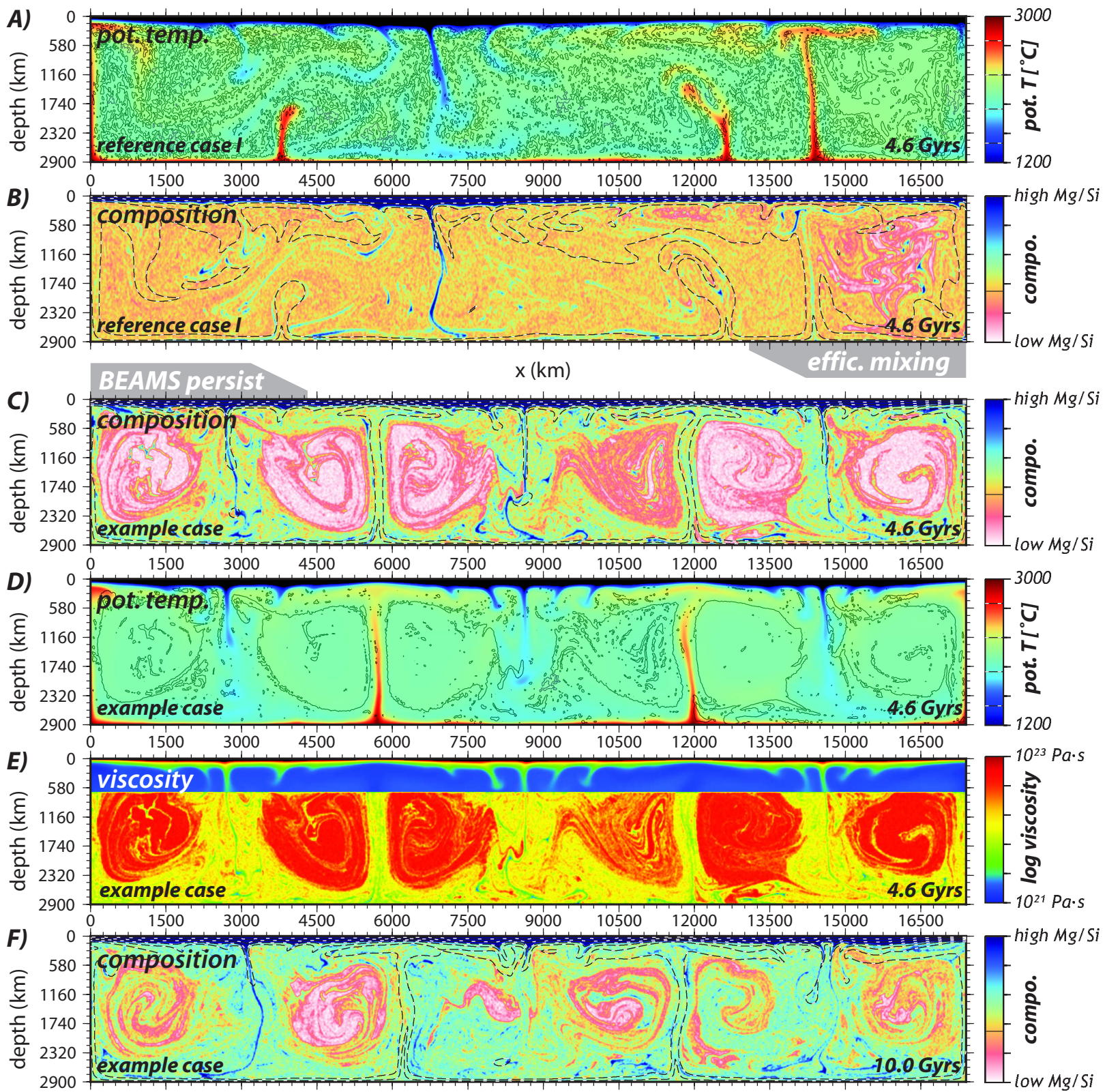

**Figure 1**

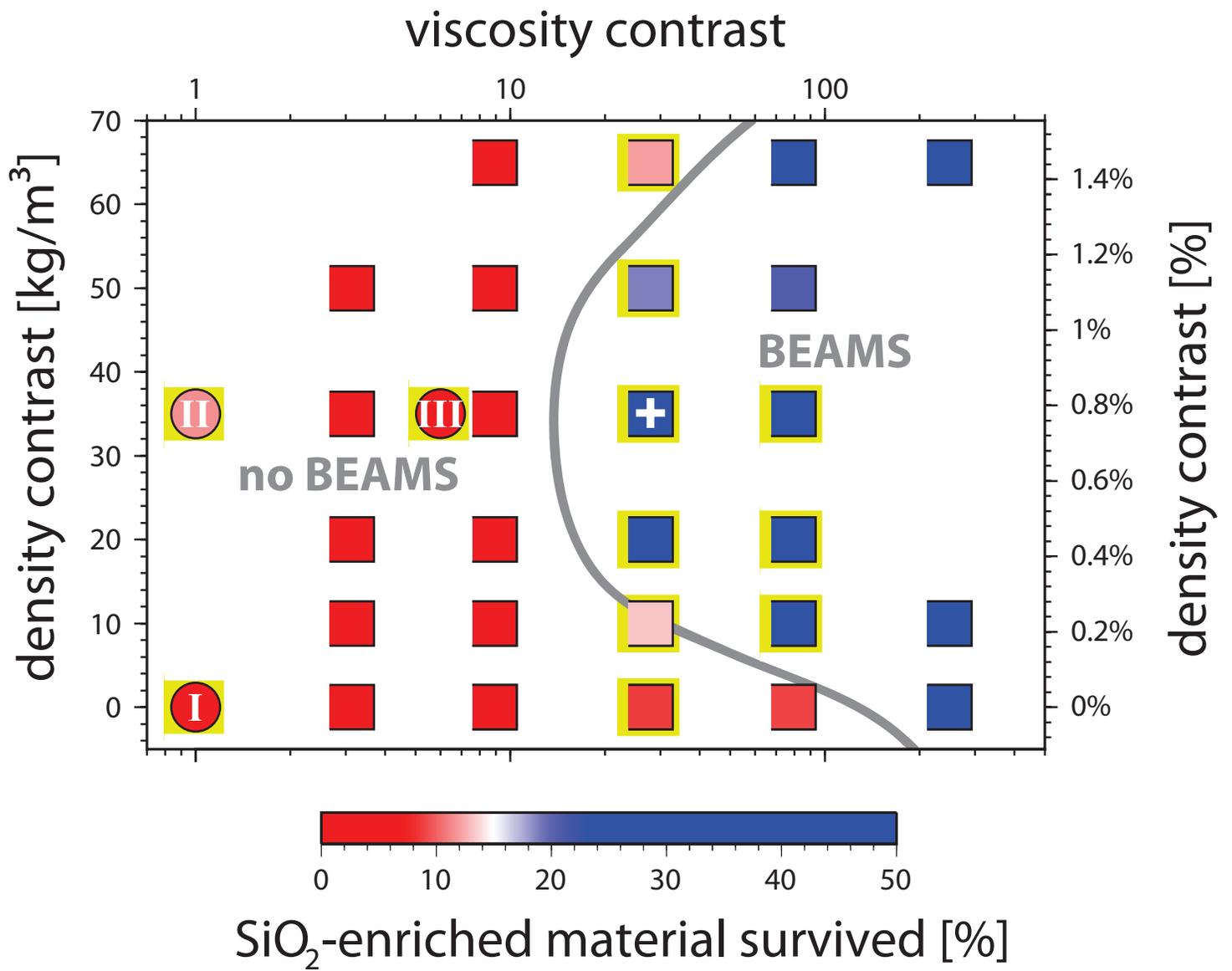

**Figure 2**

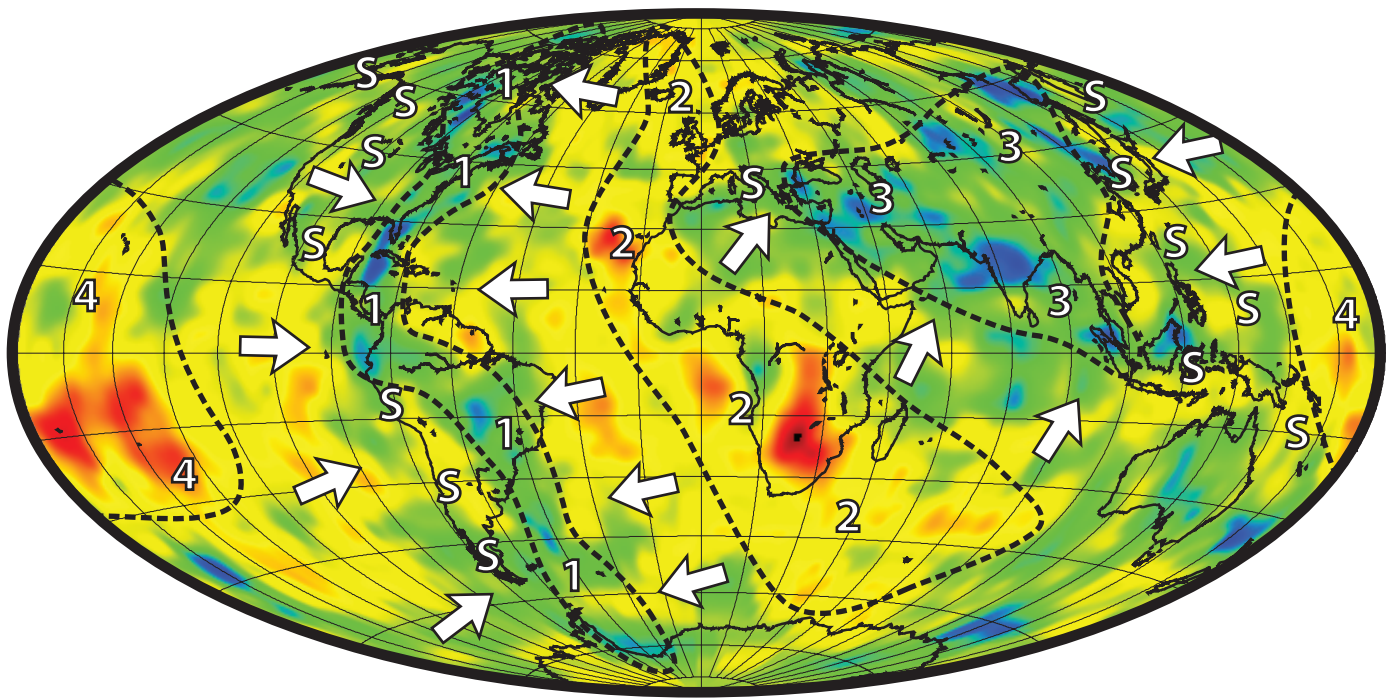
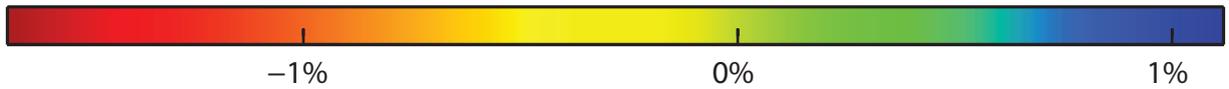

Lateral Shear Velocity Anomaly (1,000-2,200 km Depth Average)

Figure 3

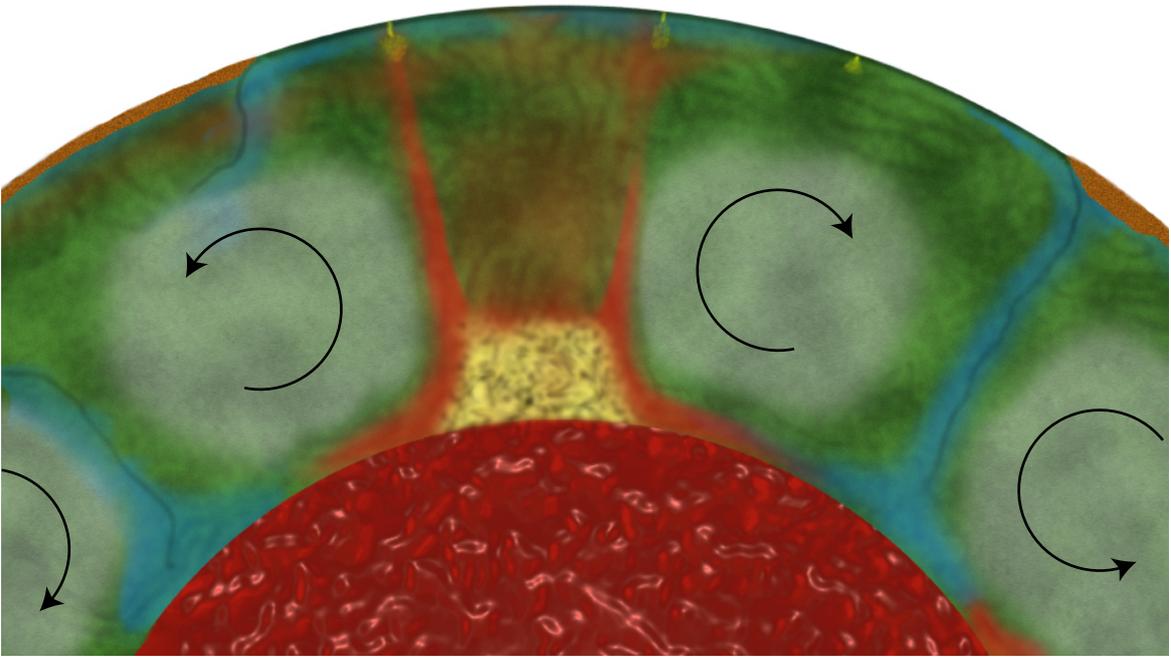

**Figure 4**

# Supplementary Information for "Persistence of Strong Silica-Enriched Domains in the Earth's Lower Mantle"

Maxim D. Ballmer[1,2], Christine Houser[1], John W. Hernlund[1], Renata M. Wentzcovitch[1,3,4] & Kei Hirose[1]

[1] Earth-Life Science Institute, Tokyo Institute of Technology, Meguro, Tokyo, Japan.
[2] Institute of Geophysics, ETH Zurich, Zurich, Switzerland.
[3] Department of Applied Physics and Applied Mathematics, Columbia University, New York, USA.
[4] Department of Earth and Environmental Sciences, Columbia University, Lamont-Doherty Earth Observatory, Palisades NY, USA.

## 1  Influence of composition on viscosity and buoyancy

In this study, we considered the influence of Mg/Si variations on viscosity and density, two key parameters governing mantle flow. The lower mantle is thought to be dominantly composed of the rheologically strong (Mg,Fe)SiO$_3$-bridgmanite (Br) phase and the relatively weak (Mg,Fe)O-ferropericlase (Fp) phase. The simplest model for deformation of a rock containing two phases is the "load bearing framework" (LBF), which is valid when strain is equally partitioned between weak grains and strong grains. The LBF approximation yields a simple linear dependence of rock viscosity upon volume fraction of one of the phases[1]. However, in rocks containing a mixture of rheologically weak and strong phases, stain is thought to be inequally partitioned. Such aggregates can hence exhibit much more dramatic variations in deformation strength and style than predicted by the LBF model, particularly if the ratio of viscosity between the two phases exceeds one order of magnitude. For large viscosity contrasts, the weak phase may become stretched and smeared between strong grains upon deformation, and if it is abundant enough to form an inter-connected network, the weaker phase may dominate the rheology of the aggregate even at modest volume fractions. As Fp is much weaker than Br, the non-linear "inter-connected weak layers" (IWL) model[1] is more appropriate, and is often invoked in viscosity estimates of lower-mantle rocks[2–4] (Suppl. Figure S5).

In the IWL model, the largest variations in viscosity occur for modal abundances of the weak phase between 0% and ∼30%. At model abundances of ∼30% and beyond, only minor variations occur, because rock viscosity is ever dominated by that of the weak phase in this range[1] (Suppl. Figure S5). The full variation in viscosity thus occurs over a range that is similar to realistic variations in Fp content in the Earth's lower mantle, based on present uncertainties in terms of the Mg/Si ratio. Therefore, the total viscosity contrast between putative $SiO_2$-enriched domains in the lower mantle and pyrolitic materials could be almost as large as the viscosity contrast between the strong and weak phases, i.e., up to three orders of magnitude.

Nevertheless, many uncertainties persist, for example in terms of the exact viscosity contrast between Br and Fp, dominant deformation mechanisms[3–5], as well as the amplitude of Mg/Si variations in the lower mantle. Additionally, the IWL model itself may be too simplistic to capture the full complexity of deformation behavior at high pressures and temperatures over geological time scales. Rock deformation is influenced by strain history, fabric, grain growth and dynamic recrystallization. Also, the viscosity contrast between Fp and Br is thought to vary as a function of depth[3,6]. In the light of these uncertainties, we explored the effects of the viscosity contrast between $SiO_2$-enriched rock and pyrolitic rocks, a free parameter in our geodynamic models. We find that modest and well-realistic viscosity contrasts of $\sim 30\times$ are sufficient for a shift in the regime of mixing of the Earth's mantle to stabilize BEAMS over the age of the Earth (see Figure 2 in the main text).

The densities of Fp and Br are similar, but not identical, due to differences in crystallographic structure. Br is slightly denser than Fp. Thus, a variation in Mg/Si also leads to a slight variation in density of the whole rock in the lower mantle. According to the calculations described in the method section, the density difference between Br and Fp steadily decreases from around 1.8% at the top of the lower mantle to around 0.8% at the base of the mantle for the same Fe/(Mg+Fe). The decrease in density contrast from 1.8% to 0.8% with increasing pressure is due to differences in bulk moduli between Br and Fp. Thus a variation of $\sim 20\%$ in Fp modal abundance, such as between pyrolite and perovskitite, naturally gives rise to a contrast of 0.16$\sim$0.36% in total rock density. This change in density (i.e., before any additional changes due to variations in Fe/(Mg+Fe)) overlaps with the ideal range of density contrasts for BEAMS persistence of 0.2$\sim$1.2% as are predicted by our geodynamic models (see Figure 2 in the main text).

## 2 Analysis of numerical-model predictions

In our suite of geodynamic models with variable density and viscosity contrasts, we observe two different regimes of mantle convection: (A) efficient mantle mixing and (B) long-term preservation of BEAMS (see main text). For details of the numerical approach, see method section. In order to distinguish between the two regimes, we analyzed the distribution of materials after 4.6 Gyrs in the model domain. Each finite element contains a mixture of materials (or material tracers), with compositional index ranging from zero to one (0: $SiO_2$-poor; 0.9-1.0: $SiO_2$-rich materials). Histograms of compositional distributions after 4.6 Gyrs model time (Suppl. Figure S6) reveal that only a subset of models display bimodal compositional distributions with significant preservation of $SiO_2$-rich material. These are the cases in which BEAMS are manifested. All other cases display unimodal distributions with one peak near compositional index 0.5.

In order to quantitatively discriminate between regimes, we evaluated the fraction of primordial $SiO_2$-rich material that has been preserved over 4.6 Gyrs of mantle convection and mixing (see Figure 2 in the main text). This preservation fraction is calculated as the number of finite elements (i.e., anywhere in the model box) with compositional index $\geq 0.84$ divided by the number of elements of the lower mantle. For example, it would be 100% if the lower mantle was entirely made up of $SiO_2$-enriched materials with compositional index $\geq 0.84$, and the upper mantle entirely of material with compositional index $<0.84$, as is the case for the initial condition.

We defined the boundary between regimes A and B at preservation fractions of ~15%, consistent with visual analysis of model results. In any case, fractions of primordial $SiO_2$-enriched material preserved after 4.6 Gyrs usually range far below 15% for regime A, and between 20% and 40% for regime B (see Suppl. Table 3). Accordingly BEAMS are predicted to make up 20%–40% of the lower mantle, or about 13%–26% of the entire mantle at the present day. Note however that these values would be lower in the 3D spherical-shell Earth's mantle, simply due to geometry. The lower mantle makes up a smaller fraction of the whole mantle in spherical geometry than in the modelled Cartesian geometry (also see below).

Figure 2 in the main text shows the boundary between regimes A (efficient mixing) and B (preservation of BEAMS) within the parameter space of variable $\Delta\rho$ and $\Phi$. One complication involves that each of the model cases displays a somewhat distinct effective-viscosity profile (through time), mostly as a function of $\Phi$. As the relevant viscosity profile controls convective vigor, and hence strongly affects mixing efficiency, the exact relevant location of the regime boundary within the parameter space remains somewhat uncertain. We quantify convective vigors by reporting the characteristic convective heat flux, or Nusselt number $Nu$, at ~4.6 Gyr (i.e., as an average over model times 4.1–5.1 Gyrs). Note that $Nu$ strictly measures the non-dimensional convective heat flux, but should be directly related to convective vigor[7], at least at given thermal-boundary-layer thicknesses, which remain virtually constant across all cases modeled (see Suppl. Table S3).

To avoid comparing cases with strongly variable $Nu$, we performed three additional cases with low $\Phi$ (see circles in Figure 2 of the main text). In these "reference" cases, an additional viscosity jump at 660 km depth of $\lambda > 0$ is imposed in order to obtain effective-viscosity profiles and convective vigors similar to those of higher-$\Phi$ models. This similarity is confirmed by comparing *Nu* between cases. For example, in Figure 2 of the main text, cases with similar *Nu* (i.e., in the range of 10–11) are highlighted yellow. Any boundary between regimes A and B based on these cases alone is similar to the boundary shown (grey line; based on all cases). This similarity confirms that our main conclusions in terms of numerical-model analysis (see below and main text) are robust.

Finally, individual analysis of reference cases shows that a viscosity jump in the mid-mantle alone is insufficient to promote large-scale preservation of distinct lower mantle domains. For example, reference case I with $\lambda = 8$ but without any effects of composition on density or viscosity (i.e., $\Phi = 1$ and $\Delta\rho = 0$ kg/m$^3$) displays efficient mantle mixing and a unimodal distribution of composition for *Nu*=10.2. Reference cases II and III confirm that small $\Phi$=6 and/or moderate $\Delta\rho$=35 kg/m$^3$ are also insufficient to avoid efficient mixing at *Nu* of ∼10.5 (see Suppl. Table 3). Note that the $\lambda$ imposed in the reference cases are chosen to tune Nusselt number to 10≤*Nu*≤11. Thereby, the $\lambda$ imposed in the reference cases remain lower bounds[8,9], mostly because we took the simplfied assumption of $\lambda = 1$ in all other cases. The latter choice implies that viscosities in the upper mantle are upper bounds, taken to limit computational costs.

We stress that all our cases with 10≤*Nu*≤11 have similar and overall realistic properties of convection, at least in the lower mantle. Maximum lower-mantle velocities in these cases (highlighted in Figure 2 of the main text) are on the order of ∼2 cm/yr, consistent with inferred slab-sinking speeds[10]. Lower-mantle viscosities are also in the realistic range[8] (see Figure 1e in the main text). Finally, dimensionalizing the above-mentioned Nu yields core-cooling-related heat fluxes of ∼30 mW/m$^2$, which are similar to those observed (∼65 mW/m$^2$) as long as radioactive heating accounts for another ∼35 mW/m$^2$ on Earth.

We conclude that compositional-viscosity contrasts of $\Phi > 20$ are essential for large-scale preservation of BEAMS (see Figure 2 of the main text). The preservation of BEAMS is an attractive scenario to address the survival of primordial geochemical reservoirs in the mantle[11,12]. An alternative scenario involves that primordial material may be preserved as piles at the base of the mantle due to the effects of intrinsic density contrasts $\Delta\rho$ of ∼3% alone[13–15]. Intrinsically-dense piles, or large low shear-velocity provinces (LLSVP)[16,17], and intrinsically-strong BEAMS may indeed be manifested together in the present-day lower mantle (see Figure 4 of the main text).

## 3   Manifestation of BEAMS in the present-day Earth's mantle

This study explores the formation of BEAMS in a 2D-Cartesian geometry, which is the simplest and most computationally efficient way to explore the essential elements of the dynamics within an extensive parameter space. The organization of convection in the BEAMS regime involves encapsulation of high-viscosity material in the core of convection cells, with low-viscosity material circulating around high-viscosity material within conduits. Such an arrangement minimizes viscous dissipation because it focuses deformation in low-viscosity materials through kinematic strain localization, and minimizes deformation in high-viscosity materials[18]. We suggest that BEAMS in the 3D spherical-shell mantle would likewise assume planforms that minimize internal deformation of high-viscosity materials. Accordingly, we expect these planforms to involve roll segments, or toroid-shaped geometries. By "segments" we refer to rolls of finite axial extent, while toroids would be donut-shaped features.

The details of BEAMS shapes hinge on whether BEAMS actually rotate or mostly remain stationary. The actual extent of BEAMS rotation should depend on the rheological coupling between BEAMS and the pyrolitic mantle that circulates around BEAMS. For non-linear rheology[4], coupling and hence BEAMS rotation is expected to be weaker than predicted by the models. Also note that inside-out rotation of donut-shaped BEAMS in the 3D spherical mantle may be strongly inhibited.

It is also important to note that the fraction of mantle occupied by BEAMS is exaggerated in our 2D models, and will be smaller in the 3D-spherical Earth due to geometrical considerations. For example, the relative lower-mantle volume alone is smaller in 3D-spherical than in 2D-Cartesion geometry. If we consider BEAMS extending from 900∼2,300 km depth as roughly roll-shaped structures (in this example ∼1,500 km in diameter), then a single 10,000 km long roll spanning ∼120 degrees of arc through the lower mantle would occupy ∼2% of the mantle's volume. Guided by our preliminary map (see Figure 3 in the main text, and discussion below), BEAMS might plausibly be assembled from 40,000-60,000 km total length of roll-like structures, hence occupying 8%∼12% volume of the mantle. The mass fraction of these BEAMS will be slightly higher owing to the elevated density of the lower mantle, giving roughly 10%∼15% of the mantle.

The current geometrical manifestation of BEAMS in the present-day lower mantle is not obvious from seismic-tomography models. The difficulty of imaging BEAMS (if they exist) is related to the small expected seismic contrasts between BEAMS and downwelling slabs (Suppl. Figure S7; also see below). Figure 3 in the main text shows an attempt to map BEAMS as well as upwelling/downwelling conduits. This attempt is guided by radially-averaged seismic shear velocity variations in the depth range of 1,000-2,200 km depth. This depth range should represent radial structure of the mid mantle by minimizing effects of the transition zone and the LLSVPs[19,20]. If we assume that conduits between the BEAMS are radially coherent features, then depth-averaging should amplify these features. As an additional guide, the location of stagnant slabs are labeled ("S")[21,22], considering that slabs may stagnate because they encounter high-viscosity BEAMS (see main text). Indeed, slab stagnation generally occurs somewhere between lower-mantle downwelling conduits (i.e., deep-sinking Tethys and Farallon slabs) and upwelling centers (i.e., beneath the south-central Pacific and Africa). In any case, note that either of the above criteria may be imperfect for mapping BEAMS, since conduits could be irregular or tilted, and alternative mechanisms may lead to slab stagnation, in addition to BEAMS[23,24]. Nevertheless, our estimate of BEAMS locations and geometry is consistent with cluster analysis of seismic-tomography models[25], given that the retrieved "neutral" cluster indeed contains BEAMS. Also note that first-order structural features in the mid-mantle such as low-velocity domains[20,25] (upwelling centers?) and high-velocity slabs of subducted Farallon and Tethys lithosphere[26] (downwelling conduits) are consistent across tomography models. Using this simple approach, we find that candidate upwelling and downwelling conduits assume the form of sheets or pillars, with intervening BEAMS assuming the form of finite rolls or donuts, compatible with the above dynamical arguments.

## 4 Comparing lower-mantle compositional models with PREM

In order to constrain lower-mantle composition, predictions for material properties of mantle rocks from experimental[27,28] and theoretical mineral physics[29–36] have been compared to 1D global seismic profiles (such as PREM[37]). However, such attempts have remained inconclusive, with proposed compositions ranging from perovskitite to pyrolite[38–46]. Much of this uncertainty in composition stems from an uncertainty in lower-mantle temperatures, given the trade-off between temperature and composition in terms of seismic velocities and density. For example, a 500 K shift in temperature changes the seismic velocity by about 1% in the lowermost mantle, similar to a shift in composition from pure Br to harzburgite. Therefore, a wide range of lower-mantle compositions may fit one-dimensional seismic profiles such as PREM within uncertainties for lower-mantle temperatures.

Supplementary Figure S2 demonstrates that both a homogenous-pyrolite-model and a BEAMS-model lower mantle (i.e., a mixture of 50% pure Br and 25% of each cold and warm harzburgite) yield a good fit to PREM, using a similar geotherm that is self-consistently calculated (see method section). The relevant geotherms are shown in Supplementary Figure S4. Thus, any comparisons of mineral-physics predictions with PREM are insufficient to constrain the bulk composition of the lower mantle. Note that even a pure Br (i.e., perovskitite) lower mantle can produce an acceptable fit using lower-mantle temperatures that are shifted by +500 K (Suppl. Figure S8).

## 5 Slab "invisibility" in the lower mantle

One of the unresolved issues in the tomographic imaging of subducted oceanic lithosphere is the weak slab signal in the mid-mantle (see main text) bracketed by a strong slab signal in the transition zone and the top of the lower mantle, as well as near the core-mantle boundary[47]. We find that the enhancement of the ambient mid-mantle by $SiO_2$-enriched material (such as in the BEAMS hypothesis) can explain this observation since the velocity contrast between relatively cold harzburgite and temperate bridgmanite is less than between relatively cold harzburgite and temperate pyrolite (Suppl. Figure S7). Another effect involves the reduction of sensitivity of bulk modulus to temperature in the mid-mantle owing to the spin transition of iron in Fp[40], but note that this effect is already taken into account in computing the curves for this figure (see also method section). Thus, the weakening slabs signal in the mid-mantle can be better explained in the context of the BEAMS model than in that of the pyrolite model.

|             | SiO$_2$        | MgO            | FeO         | CaO         |
| ----------- | -------------- | -------------- | ----------- | ----------- |
| pyrolite    | 48.01 (40.32)  | 40.89 (51.20)  | 7.67 (5.39) | 3.43 (3.09) |
| harzburgite | 44.16 (36.60)  | 46.17 (57.04)  | 8.76 (6.07) | 0.91 (0.81) |
| bridgmanite | 57.84 (50.15)  | 31.86 (41.18)  | 7.67 (5.51) | 3.43 (3.19) |

**Supplementary Table** 1: Compositions used in this study in wt.-% (mol-%). The original pyrolite composition is modified from ref. [48] by equally converting the 2.2 mol-% Al$_2$O$_3$ to MgO and SiO$_2$. The pure bridgmanite composition is derived by starting with the pyrolite composition and reducing the Mg/Si ratio until there is almost no free (Mg,Fe)O. The harzburgite is modified from ref. [34] by dividing the original 0.53 mol-% of Al$_2$O$_3$ equally between MgO and SiO$_2$.

| Parameter | Symbol | Value |
|---|---|---|
| box height | $z_{box}$ | 2,900 km |
| box width | $x_{box}$ | 17,400 km |
| CMB temperature | $T_{CMB}$ | 3,000 °C |
| Rayleigh number | $Ra$ | $2.68889 \cdot 10^7$ |
| effective upper-mantle viscosity | $\rho_m$ | $1.2 \cdot 10^{21}$ Pa·s |
| mantle reference density | $\rho_m$ | 4,500 kg/m$^3$ |
| activation energy | $E^*$ | 35.662 kJ/mol |
| thermal diffusivity | $\kappa$ | $2.5 \cdot 10^{-6}$ m$^2$/s |
| thermal expansivity | $\alpha$ | $a$ |
| **viscosity contrast between materials** | **$\Phi$** | **1 – 249.1** |
| **density contrast between materials** | **$\Delta\rho$** | **0 – 65 kg/m$^3$** |
| **viscosity jump at 660 km depth** | **$\lambda$** | **1 – 8** |
| **non-dimensional internal heating** | **$Q$** | **0 – 1** |

**Supplementary Table** 2: Parameters used in geodynamic models. The bottom four rows (bold) report the free parameters of the study (see Suppl. Table 3). ($^a$) For description of depth-dependent parameter $\alpha$, see ref. [24]. *Ra* is calculated from $\rho_m$, which is valid for the upper mantle at potential temperatures of T$_{CMB}$.

| Case | Φ | $\Delta\rho$ (kg/m$^3$) | $\lambda$ | $Q$ | Nu | SiO$_2$-rich material preserved |
|---|---|---|---|---|---|---|
| example | 27.95 | 35 | 1 | 0 | 10.83 | 30.8% |
| reference I | 1 | 0 | 8 | 0 | 10.2 | 2.18% |
| reference II | 1 | 0 | 8 | 0 | 10.78 | 11.6% |
| reference III | 6.0 | 0 | 2.5 | 0 | 10.56 | 3.9% |
| A0 | 3.14 | 0 | 1 | 0 | 11.61 | 0.41% |
| A10 | 3.14 | 10 | 1 | 0 | 12.58 | 0.13% |
| A20 | 3.14 | 20 | 1 | 0 | 11.72 | 0.02% |
| A35 | 3.14 | 35 | 1 | 0 | 11.71 | 0.01% |
| A50 | 3.14 | 50 | 1 | 0 | 12 | 0.01% |
| B0 | 8.91 | 0 | 1 | 0 | 11.81 | 3.79% |
| B10 | 8.91 | 10 | 1 | 0 | 11.45 | 5.05% |
| B20 | 8.91 | 20 | 1 | 0 | 11.71 | 6.71% |
| B35 | 8.91 | 35 | 1 | 0 | 11.72 | 6.46% |
| B50 | 8.91 | 50 | 1 | 0 | 11.15 | 1.03% |
| B65 | 8.91 | 65 | 1 | 0 | 11.19 | 0.35% |
| C0 | 27.95 | 0 | 1 | 0 | 10.45 | 8.86% |
| C10 | 27.95 | 10 | 1 | 0 | 10.83 | 13.23% |
| C20 | 27.95 | 20 | 1 | 0 | 10.83 | 30.75% |
| C35 | 27.95 | 35 | 1 | 0 | 10.84 | 30.8% |
| C50 | 27.95 | 50 | 1 | 0 | 10.7 | 18.81% |
| C65 | 27.95 | 65 | 1 | 0 | 10.34 | 12.09% |
| D0 | 79.43 | 0 | 1 | 0 | 9.45 | 9.05% |
| D10 | 79.43 | 10 | 1 | 0 | 10.24 | 31.85% |
| D20 | 79.43 | 20 | 1 | 0 | 10.49 | 37.34% |
| D35 | 79.43 | 35 | 1 | 0 | 10.22 | 36.76% |
| D50 | 79.43 | 50 | 1 | 0 | 9.55 | 20.63% |
| D65 | 79.43 | 65 | 1 | 0 | 9.44 | 0.01% |
| E0 | 249.07 | 0 | 1 | 0 | 9.33 | 33.25% |
| E10 | 249.07 | 10 | 1 | 0 | 9.4 | 37.09% |
| E65 | 249.07 | 65 | 1 | 0 | 8.05 | 49.39% |

**Supplementary Table** 3: (previous page) List of all cases modeled. Controlling parameters (see Table S2) are given in columns 2–5. Key output variables in columns 6–7 (right side). Note that case C35 and the example case are the same case. The reported *Nu* is the average *Nu* over model times 4.1-5.1 Gyrs. The reported amounts of $SiO_2$-rich material preserved is calculated for model time 4.6 Gyrs (for details, see Suppl. Section 4).

# 6 Supplementary References

1. Takeda, Y. Flow in rocks modelled as multiphase continua: Application to polymineralic rocks. *Journal of Structural Geology* **20**, 1569–1578 (1998).

2. Yamazaki, D. & Karato, S.-I. Some mineral physics constraints on the rheology and geothermal structure of Earth's lower mantle. *Amer. Mineralogist* **86**, 385–391 (2001).

3. Marquardt, H. & Miyagi, L. Slab stagnation in the shallow lower mantle linked to an increase in mantle viscosity. *Nature Geoscience* **8**, 311–314 (2015).

4. Girard, J., Amulele, G., Farla, R., Mohiuddin, A. & Karato, S.-I. Shear deformation of bridgmanite and magnesiowüstite aggregates at lower mantle conditions. *Science* **351**, 144–147 (2016).

5. Kraych, A., Carrez, P. & Cordier, P. On dislocation glide in $MgSiO_3$ bridgmanite at high-pressure and high-temperature. *Earth and Planetary Science Letters* **452**, 60–68 (2016).

6. Piet, H. *et al.* Spin and valence dependence of iron partitioning in Earths deep mantle. *Proceedings of the National Academy of Sciences* **113**, 11127–11130 (2016).

7. Parmentier, E. M., Turcotte, D. L. & Torrance, K. E. Studies of finite amplitude non-Newtonian thermal convection with application to convection in the Earth's mantle. *Journal of Geophysical Research* **81**, 1839–1846 (1976).

8. Forte, A. & Mitrovica, J. Deep-mantle high-viscosity flow and thermochemical structure inferred from seismic and geodynamic data. *Nature* **410**, 1049–1056 (2001).

9. Rudolph, M., Lekic, V. & Lithgow-Bertelloni, C. Viscosity jump in Earth's mid-mantle. *Science* **350**, 1349–1352 (2015).

10. Van Der Meer, D. G., Spakman, W., Van Hinsbergen, D. J., Amaru, M. L. & Torsvik, T. H. Towards absolute plate motions constrained by lower-mantle slab remnants. *Nature Geoscience* **3**, 36–40 (2010).

11. Mukhopadhyay, S. Early differentiation and volatile accretion recorded in deep-mantle neon and xenon. *Nature* **486**, 101–104 (2012).

12. Rizo, H. *et al.* Preservation of Earth-forming events in the tungsten isotopic composition of modern flood basalts. *Science* **352**, 809–812 (2016).

13. Deschamps, F. & Tackley, P. J. Searching for models of thermo-chemical convection that explain probabilistic tomography. IIInfluence of physical and compositional parameters. *Physics of the Earth and Planetary Interiors* **176**, 1–18 (2009).

14. Li, M., McNamara, A. K. & Garnero, E. J. Chemical complexity of hotspots caused by cycling oceanic crust through mantle reservoirs. *Nature Geoscience* **7**, 366–370 (2014).

15. Nakagawa, T. & Tackley, P. J. Influence of combined primordial layering and recycled MORB on the coupled thermal evolution of Earth's mantle and core. *Geochemistry, Geophysics, Geosystems* **15**, 619–633 (2014).

16. Garnero, E. J. & McNamara, A. K. Structure and dynamics of Earth's lower mantle. *Science* **320**, 626–628 (2008).

17. Deschamps, F., Cobden, L. & Tackley, P. J. The primitive nature of large low shear-wave velocity provinces. *Earth and Planetary Science Letters* **349**, 198–208 (2012).

18. Busse, F. Multiple solutions for convection in a two component fluid. *Geophys. Res. Lett.* **9**, 519–522 (1982).

19. Hernlund, J. & Houser, C. On the distribution of seismic velocities in Earth's deep mantle. *Earth Planet. Sci. Lett.* **265**, 423–437 (2008).

20. Lekic, V., Cottaar, S., Dziewonski, A. & Romanowicz, B. Cluster analysis of global lower mantle tomography: A new class of structure and implications for chemical heterogeneity. *Earth Planet. Sci. Lett.* **357-358**, 68–77 (2012).

21. Fukao, Y., Widiyantoro, S. & Obayashi, M. Stagnant slabs in the upper and lower mantle transition region. *Rev. Geophys.* **39**, 291–323 (2001).

22. Fukao, Y. & Obayashi, M. Subducted slabs stagnant above, penetrating through, and trapped below the 660 km discontinuity. *J. Geophys. Res.* **118**, 5920–5938 (2013).

23. King, S., Frost, D. & Rubie, D. Why cold slabs stagnate in the transition zone. *Geology* **43**, 231–234 (2015).

24. Ballmer, M. D., Schmerr, N. C., Nakagawa, T. & Ritsema, J. Compositional mantle layering revealed by slab stagnation at ∼1000-km depth. *Science Advances* **1**, doi:10.1126/sciadv.1500815 (2015).

25. Cottaar, S. & Lekic, V. Morphology of seismically slow lower-mantle structures. *Geophysical Journal International* **207**, 1122–1136 (2016).

26. Grand, S. P. Mantle shear-wave tomography and the fate of subducted slabs. *Philosophical Transactions of the Royal Society of London A: Mathematical, Physical and Engineering Sciences* **360**, 2475–2491 (2002).

27. Jackson, I. Elasticity, composition, and temperature of the Earth's lower mantle: A reappraisal. *Geophys. J. Int.* **134**, 291–311 (1998).

28. Murakami, M., Ohishi, Y., Hirao, N. & Hirose, K. A perovskitic lower mantle inferred from high–pressure, high–temperature sound velocity data. *Nature* **485**, 90–95 (2012).

29. Valencia-Dardona, J., Shukla, G., Wu., Z., Yuen, D. & Wentzcovitch, R. Impact of spin crossover in ferropericlase on the lower mantle adiabat. *Geophys. Res. Lett.* in review (2016).

30. da Silva, C., Wentzcovitch, R., Patel, A., Price, G. & Karato, S. The composition and geotherm of the lower mantle: Constraints from the calculated elasticity of silicate perovskite. *Phys. Earth Planet. Int.* **118**, 103–109 (2000).

31. Karki, B., Wentzcovitch, R., de Gironcoli, S. & Baroni, S. First principles thermoelasticity of $MgSiO_3$-perovskite: Consequences for the inferred properties of the lower mantle. *Geophys. Res. Lett.* **28**, 2699–2702 (2001).

32. Wentzcovitch, R., Karki, B., Cococcioni, M. & de Gironcoli, S. Thermoelastic properties of of $MgSiO_3$-perovskite: Insights on the nature of the Earth s lower mantle. *Phys. Rev. Lett.* **92**, doi:10.1103/PhysRevLett.92.018501 (2004).

33. Wentzcovitch, R. *et al.* Anomalous compressibility of ferropericlase throughout the iron spin crossover. *Proc. Natl. Acad. Sc. USA* **106**, 8447–8452 (2009).

34. Xu, W., Lithgow-Bertelloni, C., Stixrude, L. & Ritsema, J. The effect of bulk composition and temperature on mantle seismic structure. *Earth Planet. Sci. Lett.* **275**, 70–79 (2008).

35. Wang, X., Tsuchiya, T. & Hase, A. Computational support for a pyrolitic lower mantle containing ferric iron. *Nature Geoscience* **8**, 556–560 (2015).

36. Wu, Z. Velocity structure and composition of the lower mantle with spin crossover in ferropericlase. *J. Geophys. Res.* **121**, 2304–2314 (2016).

37. Dziewonski, A. & Anderson, D. Preliminary reference Earth model. *Phys. Earth Planet. Inter.* **25**, 297–356 (1981).

38. Shukla, G. *et al.* Thermoelasticity of $Fe^{2+}$-bearing bridgmanite. *Geophys. Res. Lett.* **42**, 1741–1749 (2015).

39. Wu, Z., Justo, J., da Silva, C., de Gironcoli, S. & Wentzcovitch, R. M. Anomalous thermodynamic properties in ferropericlase throughout its spin crossover transition. *Physical Review B* **80**, doi:10.1103/PhysRevB.80.014409 (2009).

40. Wu, Z. & Wentzcovitch, R. Spin crossover in ferropericlase and velocity heterogeneities in the lower mantle. *Proc. Nat. Academy Sci.* **111**, 10468–10472 (2014).

41. Cobden, L. *et al.* Thermochemical interpretation of 1-D seismic data for the lower mantle: The significance of nonadiabatic thermal gradients and compositional heterogeneity. *Journal of Geophysical Research: Solid Earth* **114**.

42. Cottaar, S., Heister, T., Rose, I. & Unterborn, C. BurnMan: A lower mantle mineral physics toolkit. *Geochemistry, Geophysics, Geosystems* **15**, 1164–1179 (2014).

43. Matas, J., Bass, J., Ricard, Y., Mattern, E. & Bukowinski, M. On the bulk composition of the lower mantle: predictions and limitations from generalized inversion of radial seismic profiles. *Geophysical Journal International* **170**, 764–780 (2007).

44. Mattern, E., Matas, J., Ricard, Y. & Bass, J. Lower mantle composition and temperature from mineral physics and thermodynamic modelling. *Geophysical Journal International* **160**, 973–990 (2005).

45. Ricolleau, A. *et al.* Density profile of pyrolite under the lower mantle conditions. *Geophysical Research Letters* **36**, doi:10.1029/2008GL036759 (2009).

46. Khan, A., Connolly, J. & Taylor, S. Inversion of seismic and geodetic data for the major element chemistry and temperature of the Earth's mantle. *Journal of Geophysical Research: Solid Earth* **113**, doi:10.1029/2007JB005239 (2008).

47. Houser, C. & Williams, Q. The relative wavelengths of fast and slow velocity anomalies in the lower mantle: Contrary to the expectations of dynamics? *Phys. Earth Planet. Int.* **176**, 187–197 (2009).

48. Green, D., Hibberson, W. & Jaques, A. Petrogenesis of mid-ocean ridge basalts. In McElhinny, M. (ed.) *The Earth: Its Origin, Structure and Evolution*, 265–299 (Academic Press, 1979).

49. Brown, J. & Shankland, T. Thermodynamic parameters in the Earth as determined from seismic profiles. *Geophys. J. R. Astron. Soc.* **66**, 579–596 (1981).

# 7 Supplementary Movie Captions

**Supplementary Movie S1**: Animation of mantle compositional evolution over 15 Gyrs model time for the example case (see Suppl. Table S3 for parameters). Mantle-convection patterns are stable over at least 15 Gyrs due to the persistence of intrisically strong BEAMS. For snapshots of this animation and colorscale, see Figure 1c,f in the main text. Tickmarks are in (km).

**Supplementary Movie S2**: Animation of mantle potential temperature over 15 Gyrs model time for the example case (see Suppl. Table S3 for parameters). For a snapshot of this animation and colorscale, see Figure 1d in the main text. Tickmarks are in (km).

**Supplementary Movie S3**: Animation of mantle compositional evolution over 4.6 Gyrs model time for reference case I (see Suppl. Table S3 for parameters). Mantle-convection patterns are chaotic and yield efficient mixing. For a snapshot of this animation and colorscale, see Figure 1b in the main text. Tickmarks are in (km).

**Supplementary Movie S4**: Animation of mantle potential temperature over 4.6 Gyrs model time for the example case (see Suppl. Table S3 for parameters). For a snapshot of this animation and colorscale, see Figure 1a in the main text. Tickmarks are in (km).

# 8 Supplementary Figure Captions

**Supplementary Figure S1**: Model initial conditions for all cases with composition (colors) and temperatures (contours, spaced 450K).

**Supplementary Figure S2**: Seismic velocities and density calculated for a uniform pyrolite composition (blue) and an idealized BEAMS mantle (magenta). The idealized BEAMS mantle is composed of 50% bridgmanite (high-viscosity ambient mantle), and 25% each cold and warm harzburgite (downwellings and upwellings, respectively). Black dots show PREM values[37]. Both scenarios are practically indistinguishable; differences are smaller than seismic-model resolution.

**Supplementary Figure S3**: Calculated density as well as shear-, bulk-, and compressional velocities for bridgmanite (magenta), pyrolite (blue), and harzburgite (green) (for compositions, see Suppl. Table 1). The geotherms for each of these compositions are shown in Supplementary Figure S4. Also see method section.

**Supplementary Figure S4**: Geotherms for each of the compositions described in Suppl. Table 1. Geotherms are a self-consistent output from the ab-initio calculations (see method section). Dashed and dotted green lines show temperature profiles for warm and cold harzburgite, respectively, as have been used to calculate the average BEAMS mantle shown in Supplementary Figure S2. Likewise, the dashed magenta line is the temperature profile for the warm bridgmanite as shown in Supplementary Figure S8. Geotherms for warm/cold compositions are calculated from foot temperatures at the top of the lower mantle that are shifted by +/- 500 K. The brown line is the adiabatic geotherm from ref. [49] for reference.

**Supplementary Figure S5**: Schematic viscosity variations for a lower-mantle assemblage containing both a strong Br and weak Fp phase. The viscosities of either Br or Fp involve uncertainties, as do rheological models and lower-mantle composition. Here we juxtapose the expected viscosity of BEAMS-like material with solar-chondritic Mg/Si (orange bar) with that of pyrolitic material (blue bar) in the lower mantle. The rheology of lower-mantle materials is bracketed by the "load-bearing framework" (LBF) model and the non-linear "inter-connected weak layers" (IWL) model. That said, the IWL model (solid line) is more realistic than the LBF model (dashed line) for a two-phase mixture with large viscosity contrast[2,4], particularly in high-strain regions such as predicted by our models for pyrolitic up-/downwelling conduits. Schematic distributions of strong Br (white) and weak Fp (black) grains are shown as inset images. Note that in the IWL model the largest change in viscosity occurs as the fraction of weak phase (Fp) varies between 0% and about 30%.

**Supplementary Figure S6**: Histograms of composition after 4.6 Gyrs model time for cases with (A) variable $\Phi$ at the same $\Delta\rho$, and (B) variable $\Delta\rho$ at the same $\Phi$. Note that all reference cases are also represented in (A) as dashed lines, including reference case I with $\Delta\rho = 0$ kg/m$^3$. In addition to a sharp peak at compositional index 0 (pure pyrolite/harzburgite composition in the shallow mantle according to model assumptions), cases either display a unimodal distribution of composition throughout the model domain (with one peak at compositional index $\sim$0.5; i.e. relatively well-mixed: regime A), or a bimodal distribution (with another peak at >0.84; i.e. preservation of SiO$_2$-rich material: remime B). The white histogram in front of the grey shading shows the original distribution of lower-mantle material at time-step zero (scaled by a factor of 0.25) for comparison. The red line is the same in both panels, and marks the example case visualized in Figure 1c-f. All other cases as labeled.

**Supplementary Figure S7**: Lateral seismic velocity variations of warm and cold harzburgite relative to ambient mantle of unknown composition. We vary the composition of the ambient adiabatic mantle, to which the warm (+500 K) or cold (-500 K) harzburgite is compared to, from pyrolitic to bridgmanitic compositions. Addition of $SiO_2$ to the background ambient mantle lowers the seismic velocity contrast of cold downwelling material and amplifies the contrast of warm upwelling material. These conclusions hold even if the effects of basalt on seismic velocities in pyrolitic domains are included (not shown). For computation of thermoelastic properties, see method section.

**Supplementary Figure S8**: Calculated density as well as shear-, bulk-, and compressional velocities for warm (+500 K) bridgmanite. The geotherm associated with this calculation is the dashed magenta line in Supplementary Figure S4. For details, see method section.

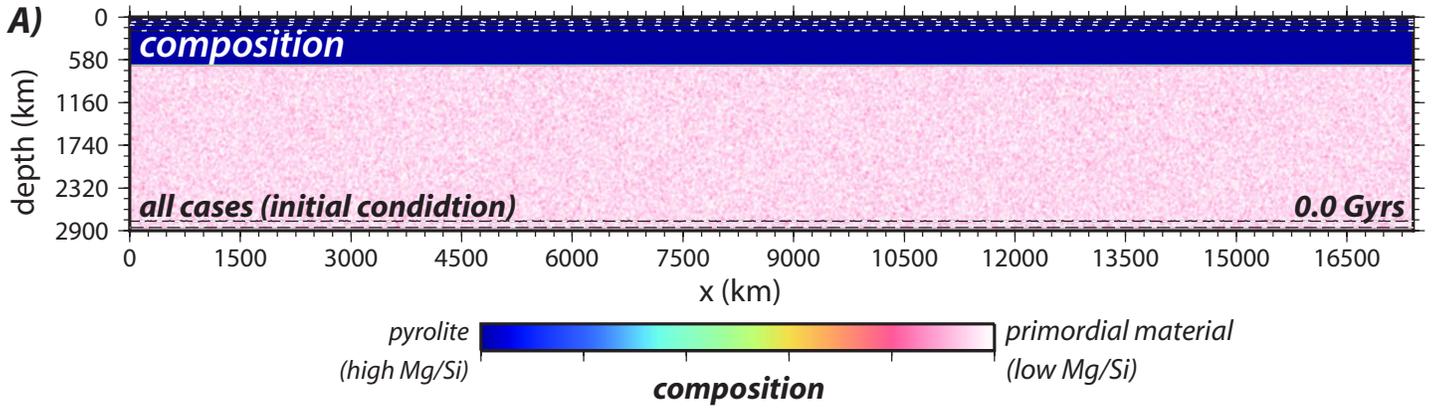

**Supplementary Figure S1**

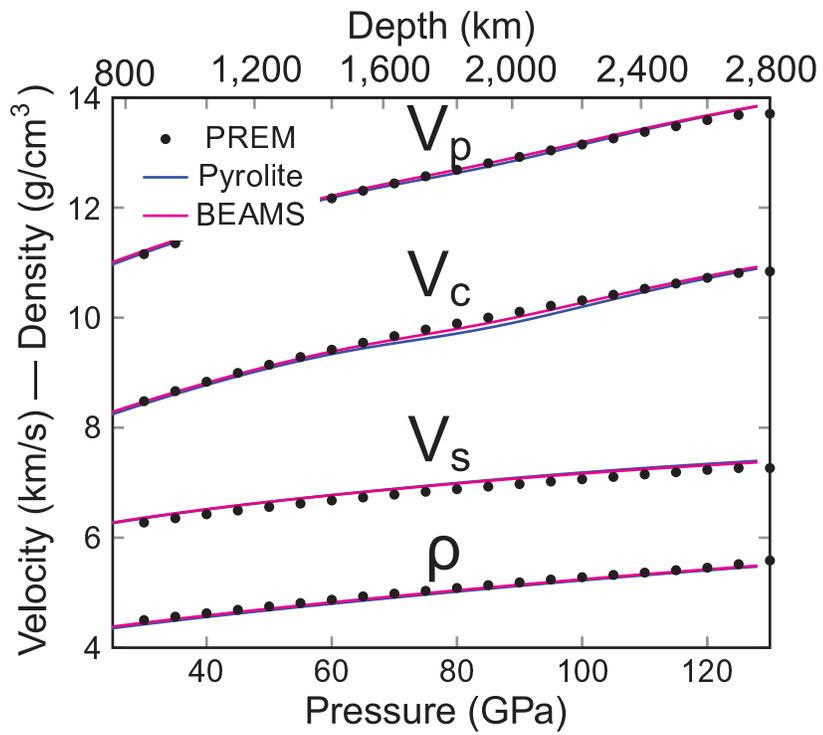

**Supplementary Figure S2**

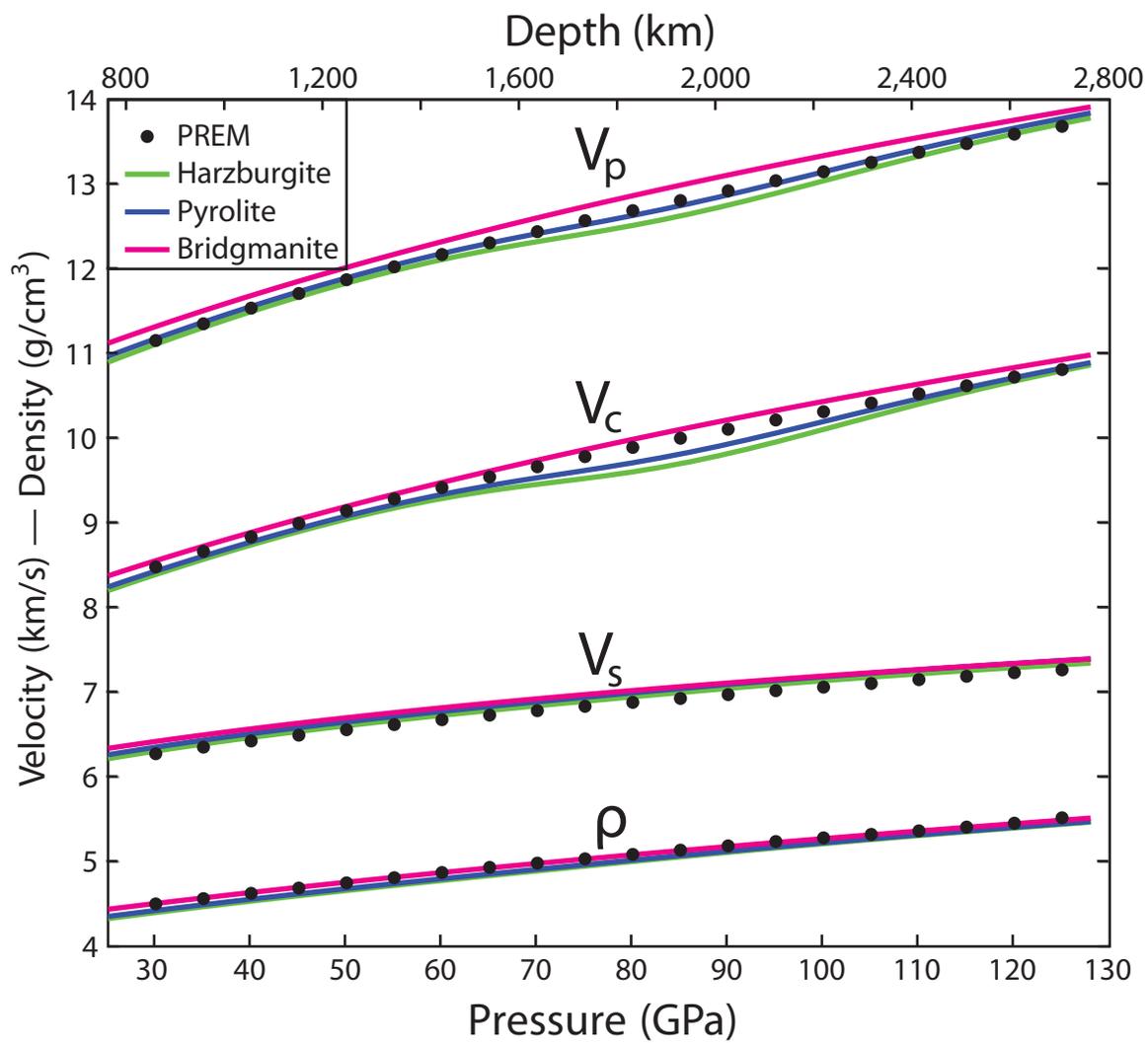

**Supplementary Figure S3**

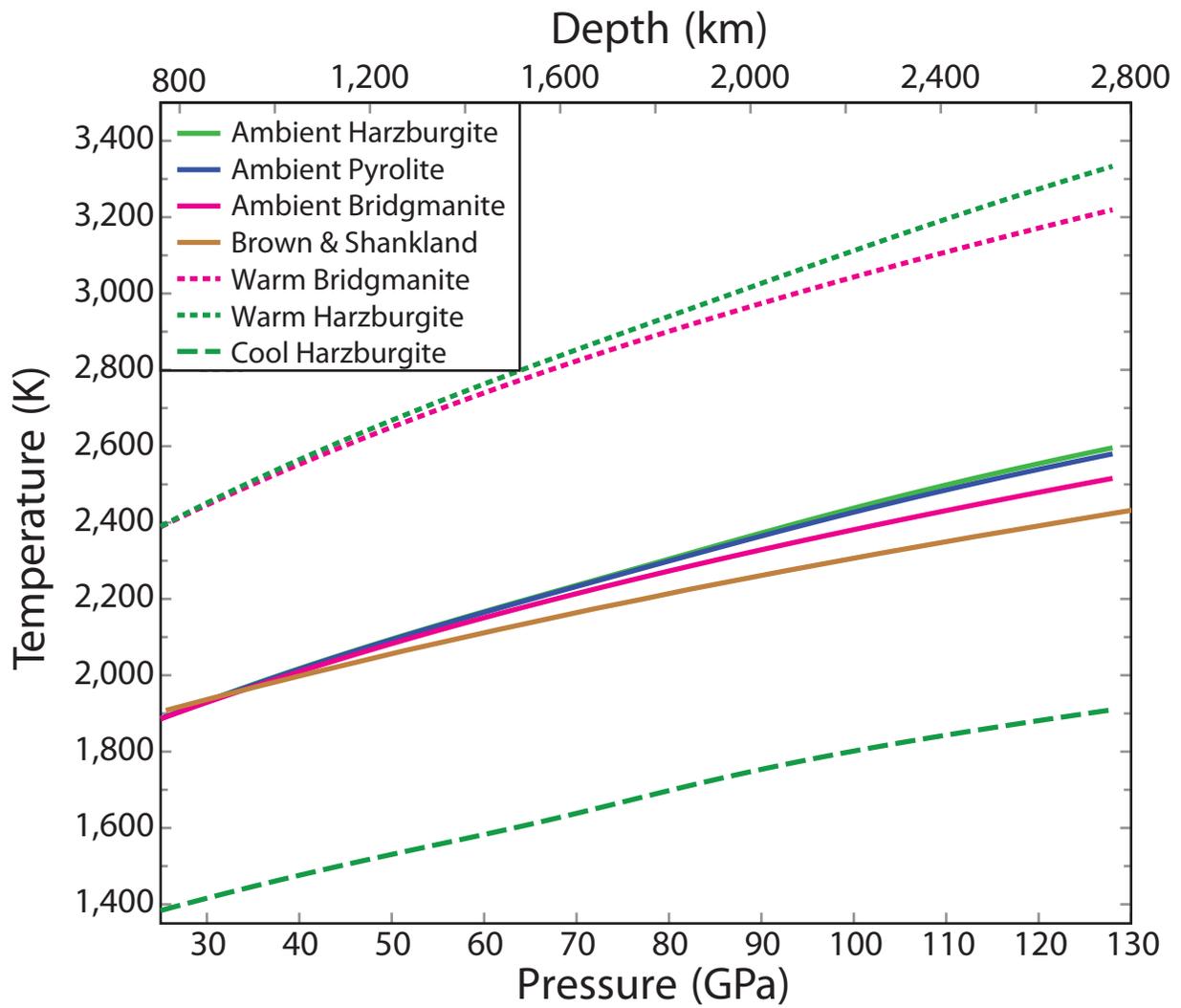

**Supplementary Figure S4**

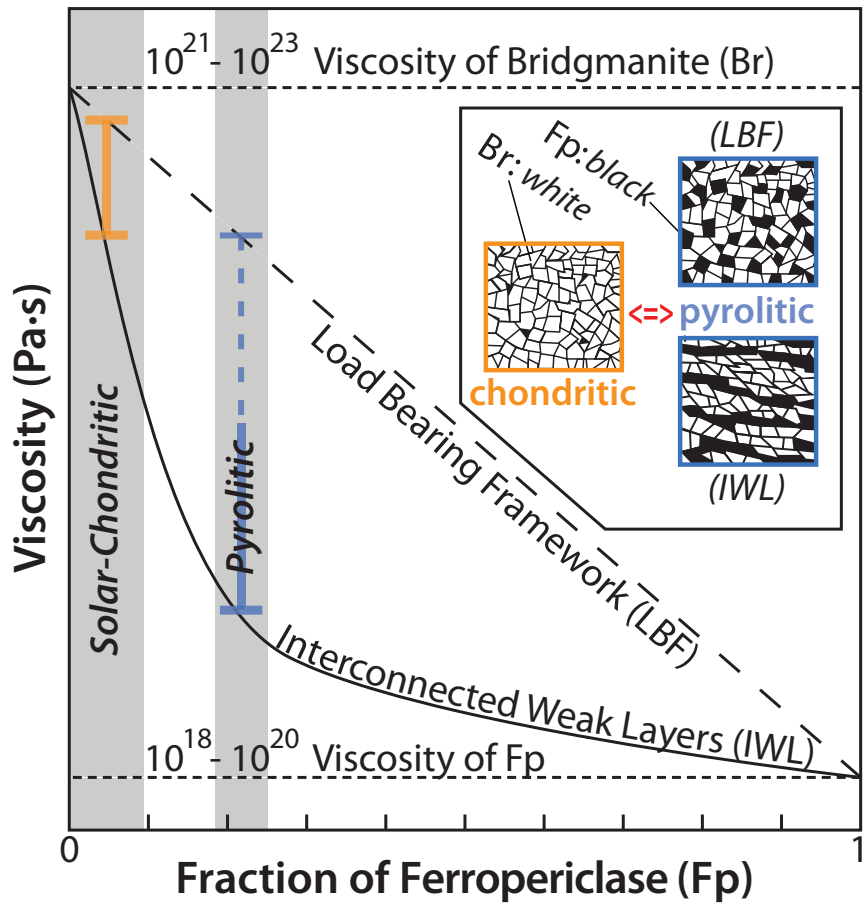

**Supplementary Figure S5**

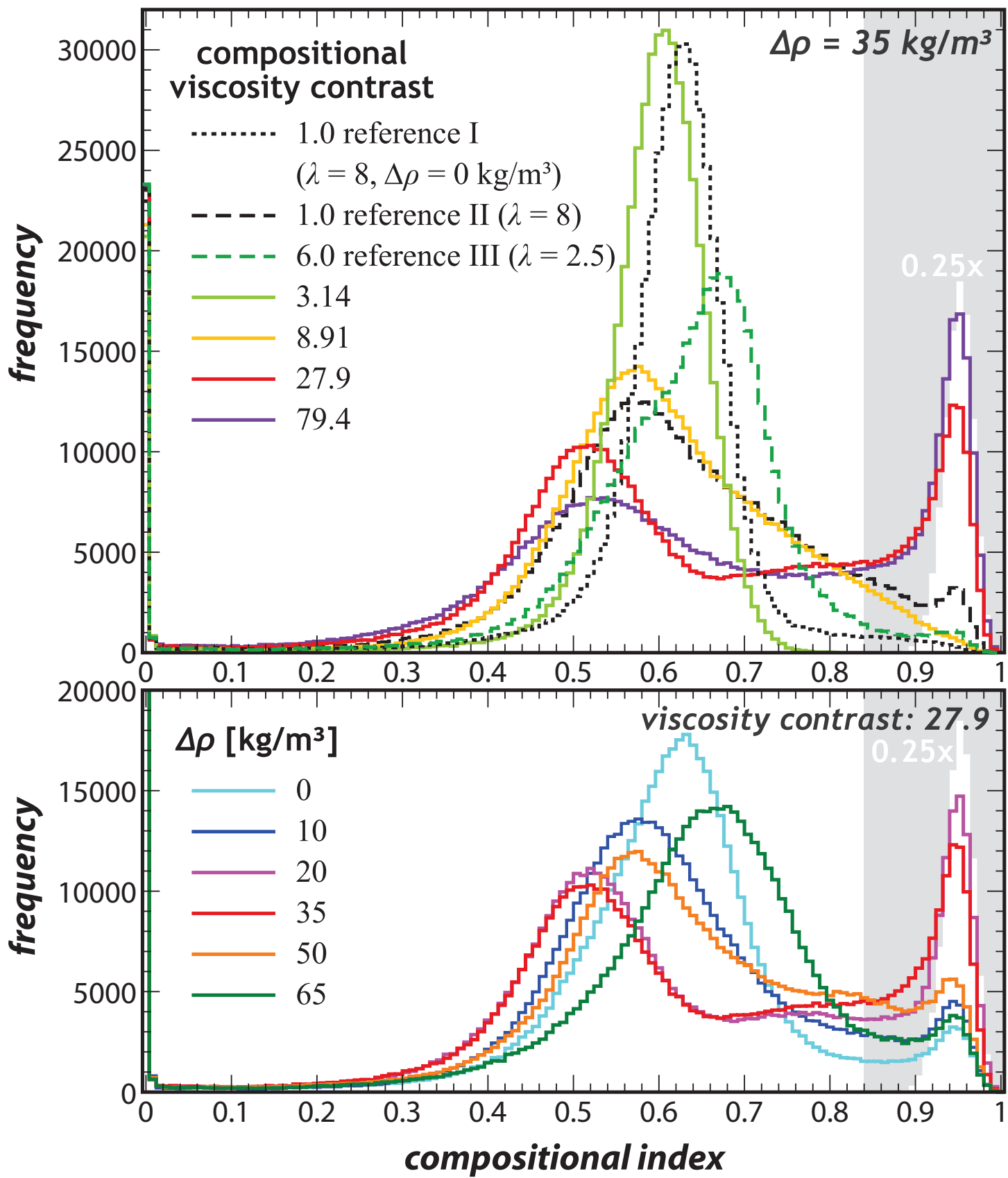

**Supplementary Figure S6**

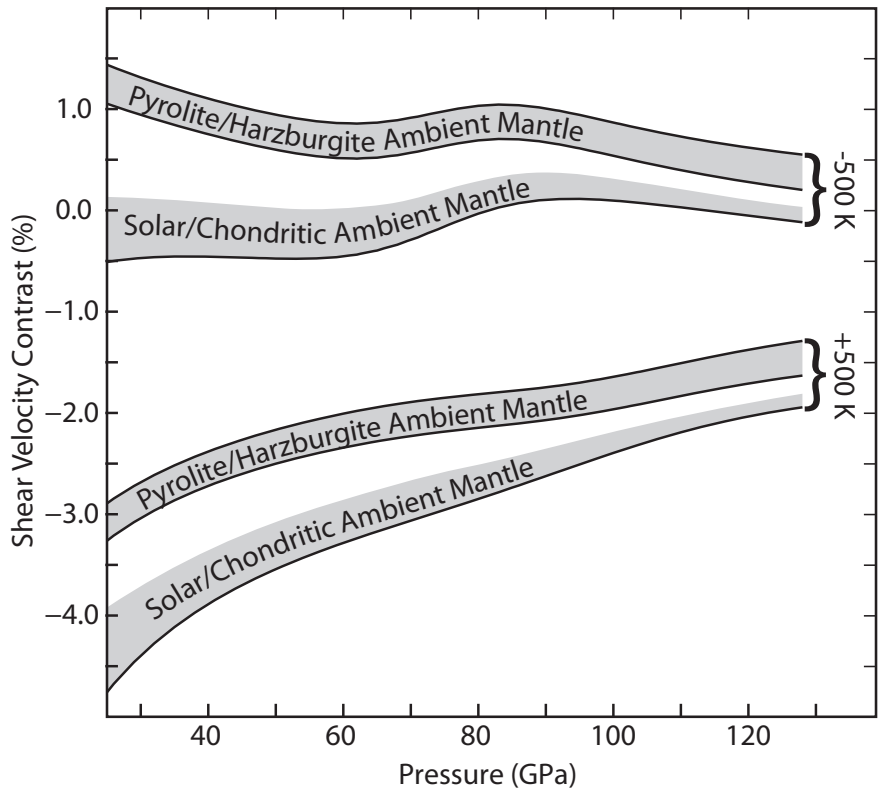

**Supplementary Figure S7**

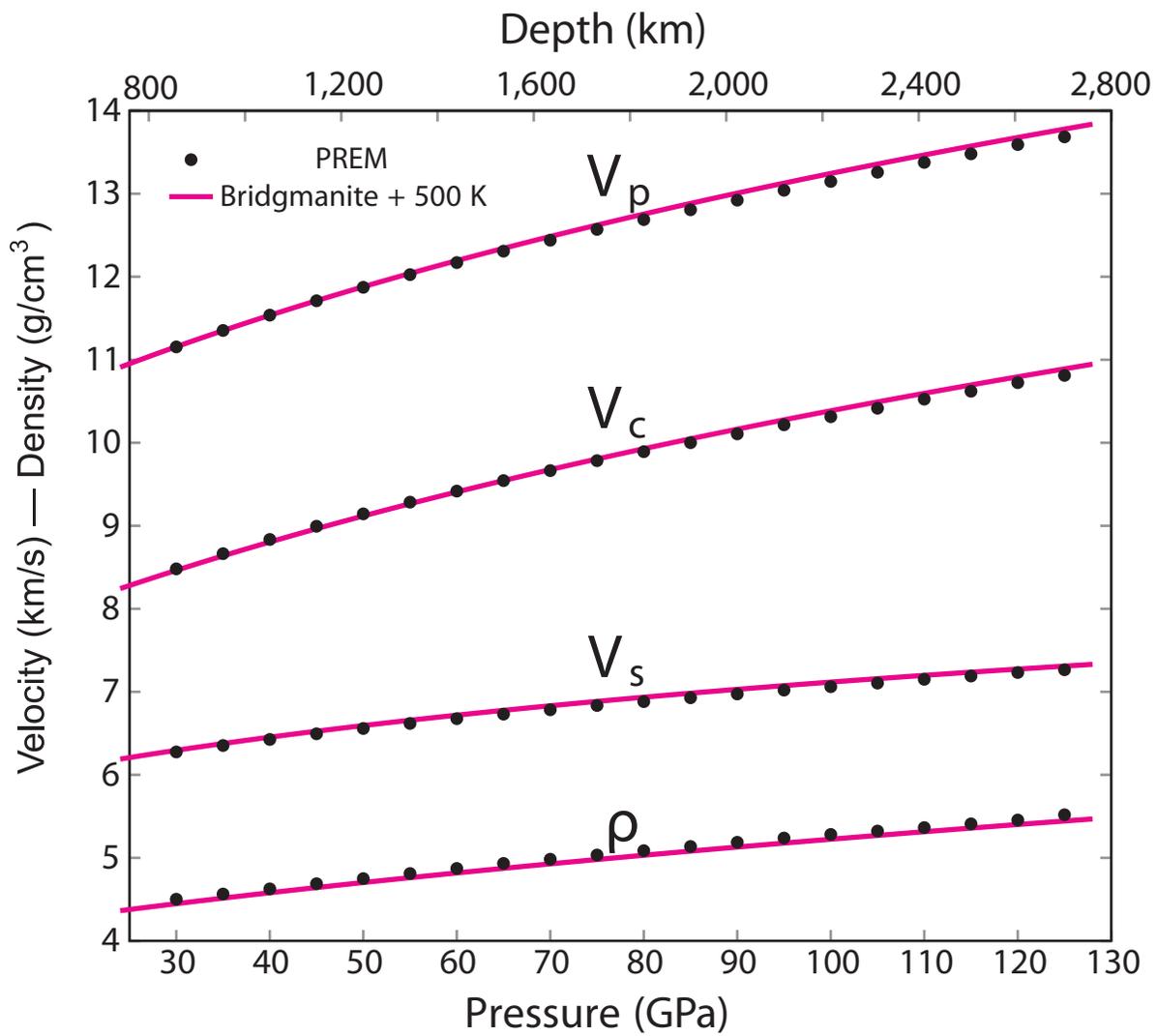

**Supplementary Figure S8**